\documentstyle[11pt,aaspp4,psfig]{article}
\def\@tightleading{0.95}
\def\baselinestretch{\@tightleading}\normalsize

\def\cg{c_{g}}
\def\cavgt{\langle c_{g}^2\rangle}
\def\gamg{{\gamma_{g}}}
\def\gamw{{\gamma_{\rm w}}}
\def\kgr{K_{g,{\rm ref}}}
\def\kavg{\avg{K_g}}
\def\kavl{\avg{K_\ell}}
\def\kpg{K_{{\rm p}g}}

\def\dlkg{\delta\ln K_{{\rm p}g}}
\def\pgref{P_{g,{\rm ref}}}
\def\plref{P_{\ell,{\rm ref}}}
\def\rhoref{\rho_{g,{\rm ref}}}
\def\mucr{\mu_{\rm cr}}

\def\s{{\rm s}}
\def\seff{\sigma_{\rm eff}}
\def\snt{\sigma_{\rm nt}}

\def\indep{\chi}
\def\kpi{K_{{\rm p}i}}
\def\kpg{K_{{\rm p}g}}
\def\dlk{\delta\ln\kpi}
\def\dlkg{\delta\ln\kpg}
\def\dll{\delta\ln\lambda}
\def\dlm{\delta\ln\mu}
\def\g{{\gamma}}
\def\gi{{\gamma_i}}
\def\gl{{\gamma_\ell}}

\def\gp{{\gamma_{\rm p}}}
\def\gpb{{\gamma_{{\rm p},B}}}
\def\gpi{{\gamma_{{\rm p}i}}}
\def\gpg{{\gamma_{{\rm p}g}}}
\def\gpl{{\gamma_{{\rm p}\ell}}}

\def\gpw{{\gamma_{\rm p,w}}}

\def\nondim{nondimensional}

\def\2D{{\rm 2D}}

\def\Mbe{M_{\rm BE}}

\def\Mw{M_{\rm w}}
\def\Mcr{M_{\rm cr}}
\def\Mcrt{M_{\rm cr,t}}
\def\mucr{\mu_{\rm cr}}
\def\rhoc{\rho_{\rm c}}
\def\rhos{\rho_{\rm s}}

\def\pip{\left(\frac{P_i}{P}\right)}
\def\plp{\left(\frac{P_\ell}{P}\right)}
\def\pgp{\left(\frac{P_{g}}{P}\right)}

\def\R{r}

\def\cavt{\avg{c^2}}
\def\chicr{\chi_{\rm cr}}
\def\cth{c_{\rm th}}
\def\gi{{\gamma_i}}
\def\gpi{{\gamma_{{\rm p}i}}}

\def\kavb{\langle K_B\rangle}

\def\kavl{\langle K_\ell\rangle}

\def\kpi{K_{{\rm p}i}}
\def\kpl{K_{{\rm p}\ell}}
\def\mbe{M_{\rm BE}}
\def\pbref{P_{B,{\rm ref}}}
\def\pth{P_{\rm th}}
\def\rhoref{\rho_{{\rm ref}}}
\def\s{{\rm s}}
\def\vref{V_{{\rm ref}}}

\def\newpage{\vfill\eject}

\def\dis{\displaystyle}

\def\beq{\begin{equation}}
\def\eeq{\end{equation}}
\def\beqa{\begin{eqnarray}}
\def\eeqa{\end{eqnarray}}
\def\bml{\begin{mathletters}}
\def\eml{\end{mathletters}}

\def\etal{{et al.}}

\def\avg#1{\langle #1\rangle}
\def\simlt{\lower.5ex\hbox{$\; \buildrel < \over \sim \;$}}
\def\simgt{\lower.5ex\hbox{$\; \buildrel > \over \sim \;$}}
\def\solar{\ifmmode _{\mathord\odot}\else $_{\mathord\odot}$\fi}
\def\Msun{\ifmmode {\rm M}_{\mathord\odot}\else $M_{\mathord\odot}$\fi}

\def\to{\ifmmode \rightarrow\else $\rightarrow$\fi}

\def\power#1{\ifmmode \times10^{#1}\else $\times~10^{#1}$\fi}
\def\none{\ifmmode ^{-1}\else $^{-1}$\fi}
\def\two{\ifmmode ^{2}\else $^{2}$\fi}
\def\ntwo{\ifmmode ^{-2}\else $^{-2}$\fi}
\def\three{\ifmmode ^{3}\else $^{3}$\fi}
\def\nthree{\ifmmode ^{-3}\else $^{-3}$\fi}
\def\four{\ifmmode ^{4}\else $^{4}$\fi}
\def\nfour{\ifmmode ^{-4}\else $^{-4}$\fi}
\def\five{\ifmmode ^{5}\else $^{5}$\fi}
\def\nfive{\ifmmode ^{-5}\else $^{-5}$\fi}

\def\cm{\ifmmode {\rm cm}\else cm\fi}
\def\m{\ifmmode {\rm m}\else m\fi}
\def\km{\ifmmode {\rm km}\else km\fi}
\def\pc{\ifmmode {\rm pc}\else pc\fi}
\def\ly{\ifmmode {\rm ly}\else ly\fi}
\def\s{\ifmmode {\rm s}\else s\fi}
\def\Hz{\ifmmode {\rm Hz}\else Hz\fi}
\def\y{\ifmmode {\rm y}\else y\fi}
\def\K{\ifmmode {\rm K}\else K\fi}
\def\ster{\ifmmode {\rm ster}\else ster\fi}
\def\erg{\ifmmode {\rm erg}\else erg\fi}
\def\dyn{\ifmmode {\rm dyn}\else dyn\fi}

\begin{document}

\title{MULTI--PRESSURE
POLYTROPES AS MODELS FOR THE STRUCTURE AND STABILITY OF MOLECULAR
CLOUDS. I. THEORY}
	
\author{Christopher F. McKee}
\affil{Departments of Physics and Astronomy,
University of California, Berkeley, CA 94720}
\and
\author{John H.~Holliman, II}
\affil{Department of Astronomy, University of California, Berkeley, CA
94720}

\rightskip=0pt
\begin{abstract}
\rightskip=0pt

	We present a theoretical formalism for 
determining the structure of
molecular clouds and the precollapse 
conditions in star--forming regions.
The model consists of a pressure-bounded, self-gravitating sphere of an
ideal gas that is supported by several distinct pressures.
Since each pressure component
is assumed to obey a polytropic law $P_i(r)\propto
\rho^\gpi$, we refer to these models as {\it
multi--pressure polytropes}.
We treat the case without rotation.
The time evolution of one of these polytropes depends additionally on 
the adiabatic index $\g_i$ of each component, which is
modified to account for the effects of any thermal coupling to the
environment of the cloud.
We derive structure equations, as well as perturbation equations for
performing a linear stability analysis.
Special attention is given to properly representing the
significant pressure components in molecular clouds:
thermal motions, static magnetic fields, and turbulence.
The fundamental approximation in our treatment is that
the effects of turbulent motions in supporting a cloud
against gravity can be approximated by a polytropic pressure
component.  In particular, we 
approximate the turbulent motions as a superposition of Alfv\'en waves.
We generalize the standard treatment of the stability of
polytropes to allow for the flow of
entropy in response to a perturbation,
as expected for the entropy associated with wave pressure.
In contrast to the pressure components within stars, 
the pressure components within interstellar clouds
are ``soft", with polytropic
indexes $\gpi\leq 4/3$ and (except for Alfv\'en waves)
adiabatic indexes $\gi\leq 4/3$.
This paper focuses on the characteristics of
adiabatic polytropes with a single pressure component
that are near the brink of gravitational instability  
as a function of $\gpi$ and $\gi$ for $\gpi\leq 4/3$.  The
properties of such polytropes are generally governed by the conditions 
at the surface. 
We obtain upper limits for the mass and size of polytropes
in terms of the 
density and sound speed at the surface.  
The mean--to--surface density and pressure drops
are limited to less than a factor 4 for $\gp\leq 1$, regardless
of the value of $\gamma$.
The central--to--surface density and pressure
drops in isentropic clouds ($\gi=\gpi$) are also limited,
but they can become quite large (as observed) in non--isentropic
clouds, which have $\gi>\gpi$.
We find that the motions associated with Alfv\'en waves
are somewhat less effective in supporting clouds than are
the kinetic motions in an isothermal gas.

\end{abstract}

\keywords{ISM: Clouds --- ISM: molecules --- ISM: 
structure --- Stars: formation}

\section{INTRODUCTION}

Star formation plays a crucial role in galactic evolution and 
in determining the structure of the interstellar medium.
In general, star-forming regions appear to be in rough hydrostatic equilibrium
since the observed supersonic line widths in giant
molecular clouds (GMCs) are due to disordered macroscopic
motions that contribute to dynamical support rather than to large-scale
infall (\cite{zuc74}).
While the large scale turbulent motions within GMCs
lead to evolution of the shape of the cloud (\cite{bal99}), they do
not lead to an overall expansion or contraction on a dynamical time
scale.  
Correspondingly, the lifetime of a GMC typically exceeds its 
free-fall timescale
($\sim10^{6.5}$~y) by about an order of magnitude (\cite{bli80}).
To a good approximation, then, most observed GMCs are in a steady
state. Structures within GMCs also appear to be in approximate equilibrium:
The density in and around a typical ammonia core within a GMC
varies roughly as $r^{-2}$ ({\it e.g.}, \cite{sne81}; \cite{ful84};
\cite{zho90}; \cite{lad91}),
a relationship that is consistent with a simple, stably stratified model.
Turner (1993) finds that the majority of cores he observes in high-latitude
cirrus clouds have characteristics that are consistent with hydrostatic
equilibrium.

	We find that spherical models accurately represent many of 
the properties of these
equilibrium structures, aside, of course, from details about the
multi-dimensional cloud geometry.
In particular, the time-honored problem of the structure and gravitational
stability of gas spheres bears upon several unresolved issues in star
formation.
What is the dependence of density and line width on size scale for a
cloud in stable hydrostatic equilibrium?
What are the initial conditions appropriate for gravitational collapse,
particularly on large scales where thermal pressure is relatively
unimportant?
How can GMCs be stable at the high mean pressures often observed?

\subsection{Molecular Cloud Structure}
\label{sec:Molecular Cloud Structure}

Because we must specify an equilibrium structure before
performing a stability analysis, our first step is to determine the
density and line width profiles of a model molecular cloud.
For a cloud in hydrostatic equilibrium, these profiles can be determined
theoretically once the nature of the pressure supporting the cloud is
specified.
We consider three sources of pressure:
thermal motions, static magnetic fields, and turbulence.

	The thermal sound speed in a given molecular cloud is roughly 
constant and
corresponds to a temperature of 10--30~K.
Thermal motions typically have significant dynamical influence only in
cloud cores (\cite{mye83}).
This pressure component may dominate support in these limited regions if
the static magnetic field is nearly straight and uniform or if the field
strength is low.

	Typically, though, static magnetic fields play a substantial role in
stabilizing molecular clouds (\cite{mou76b}; \cite{hei93}).
Magnetic support is particularly significant in regions that are forming
clusters of stars.
Since the mass of a star-forming core exceeds its thermal Jeans mass,
a clump containing multiple cores must contain many thermal Jeans masses;
the material in the clump thus owes much of its dynamical support to
the static magnetic field.

	Turbulent motions provide another source of nonthermal support
of molecular clouds.
On large scales, turbulent motions
have an energy density comparable to
that of the static magnetic field (\cite{mye88}).
These motions are supersonic, and are 
observed to be an increasing function of map size
(\cite{lar81}; \cite{cer85}; \cite{mye92}).
It was suggested some time ago that these motions are large-amplitude
MHD waves (\cite{aro75});
recent observations (\cite{mye88}) and theoretical results
(\cite{bon87}; \cite{pud90}; \cite{mck95}) are
consistent with this idea.
Alfv\'en waves, being non-compressive, are expected
to be the dominant form of MHD waves present (\cite{mck95}).
The pressure due to the waves includes both the dynamic
pressure associated with the motion of the gas and
the magnetic pressure associated with 
time-dependent magnetic field.
The importance of the wave pressure increases with
size scale for two reasons:
first, there is an increase in the range of wavenumbers 
for the MHD waves (\cite{bon87}), and second, the density
decreases with size (\cite{lar81}), which leads to an increase in
the amplitude of the waves (\cite{fat93}; \cite{mck95}).
Numerical simulations confirm that MHD waves are an important
contributor to the support of molecular clouds, but
they also indicate that the waves damp out very rapidly
(\cite{gam96}, \cite{vaz99}; but see \cite{mck99}).

	The primary pressure components that support 
molecular clouds against gravitational collapse have been identified,
and there is substantial agreement as to when each plays an important
role:
Static fields are important throughout molecular clouds,
whereas thermal pressure is important primarily in small,
dense regions and turbulent, or wave, pressure is important
on large scales.  However, there is no quantitative understanding
as to how these three sources of pressure interact to produce
the density and velocity structure that is observed in molecular
clouds.  
To date there has been only one study of the case in which
all three components are present (\cite{liz89}).
In this work, as well as in subsequent work
(Gehman et al, 1996; McLaughlin \& Pudritz 1996),
a phenomenological model is adopted for the wave pressure,
in which it
varies as the logarithm of the density
(a ``logatrope").  The disadvantage of such phenomenological models
is that it is impossible to study the stability
of the clouds in a physically motivated way.
Our intention is to develop a theoretical
framework in which the distinct pressure components can be
modeled in such a manner that both the structure and the stability
of the clouds can be determined.

\subsection{Precollapse Conditions in Star-Forming Regions}
\label{sec:Precollapse Conditions in Star-Forming Regions}

	The state of a cloud at the onset of collapse determines
the development of the density and velocity profiles
once collapse begins, the fragmentation
of the cloud, the luminosity of the protostars that form, and the final
stellar configuration.
We can illustrate the effects of the precollapse conditions by comparing
self-similar solutions for the collapse of an isothermal sphere, which
is often used to model low-mass star formation.
There is an infinite number of such solutions (Hunter 1977),
with the solution discovered by Larson (1969) and Penston (1969) and the
solution found by Shu (1977) demonstrating the ranges of velocity profiles
and densities that are possible.
The Larson-Penston (LP) solution starts from a static, uniform gas
that fills a large spherical volume.  As the gas collapses under
the influence of gravity, a rarefaction 
front propagates inward from the outer boundary at the sound speed.
When the rarefaction reaches the origin,
a condensed object forms.  At this instant, the inflow is 
spatially constant
at 3.3 times the sound speed; the density 
has the same scaling as a singular isothermal sphere (SIS),
$\rho\propto r^{-2}$, but its magnitude is 4.4 times greater.
Shu (1977) considered the other extreme, in which collapse begins from the
equilibrium state described by the SIS.
This solution is ``inside-out" in the sense that after a perturbation at the
origin, information that internal regions are evolving dynamically propagates
outward in an expansion wave.
Since the initial state is static, with $\rho\propto r^{-2}$, 
core formation begins
immediately, before a flow field develops.
Following the formation of a condensed object in both the LP solution
and the Shu solution, the region around
this object acquires power-law profiles of the form
$\rho\propto r^{-3/2}$ and $v\propto r^{-1/2}$;
however, the density and velocity at comparable points remain larger
in the LP solution.
As a result, the mass accretion rate and the accretion
luminosity are larger in the LP solution than in the Shu solution.

	Results from studies of the dynamical evolution of
collapsing  isothermal
clouds often resemble the analytic solutions described above.
An isothermal sphere in an unstable, but non-singular, equilibrium has
a density profile that is approximately flat (like the initial state of
the LP solution) in a central region, but that resembles the $r^{-2}$ form
of the SIS in an outer region.
As a result, simulations of collapse from such initial conditions
(\cite{hun77}; \cite{fie93}; \cite{fos93})
display densities and flows near their incipient condensed
objects that closely resemble those in the LP solution.
By contrast, the material in the outer part of the sphere maintains acoustic
contact while adjusting to imbalances between pressure gradients and gravity;
thus, in this region, the simulation more closely reproduces the zero
flow beyond the expansion wave in the SIS collapse solution.
If the part of the sphere that approximates
$\rho\propto r^{-2}$ prior to collapse is sufficiently large,
the accretion rate eventually approaches the constant mass
accretion rate of the SIS (\cite{fos93}).
Similar results are found for the gravitational collapse of a cloud
supported by a magnetic field undergoing ambipolar diffusion
(Safier, McKee, \& Stahler 1997).
The initial state of the cloud also affects the stability of
the collapse
with respect to fragmentation,
an essential process in the formation of star systems and clusters
(\cite{bos88}).
The collapse of a uniform-density sphere is unstable to fragmentation,
while that of an isothermal sphere 
is much less so (\cite{shu77}; \cite{sil88}).
Simulations ({\it e.g.}, \cite{bos87}) confirm that
a more centrally concentrated initial density profile
inhibits fragmentation during dynamical collapse.

	This discussion illustrates the impact of the precollapse
conditions on star formation in the idealized case of an
isothermal sphere.  In reality, 
the gas may not be isothermal, and
magnetic fields and turbulent pressure both contribute to the support
of the cloud.  One simplification remains possible
for low mass star formation, however:
it is believed to proceed quiescently,
as ambipolar diffusion gradually reduces the pressure support
due to static magnetic fields (\cite{mes56}; \cite{shu87};
\cite{mou91}).
In this case, the collapse proceeds from a
{\it critical point}---an equilibrium state on the verge of gravitational
instability.  By this criterion, neither a homogeneous sphere nor
an SIS are accurate representations of the precollapse
state, since the first is not an equilibrium and the second is
an unstable equilibrium.
Nevertheless, highly centrally concentrated initial states are possible;
calculations indicate that static magnetic fields provide a substantial
stabilizing influence both in terms of the critical mass they support
and the maximum center-to-surface density ratio that they allow
(\cite{mou76b}; Tomisaka, Ikeuchi, \& Nakamura 1988b\ [hereafter TIN];
\cite{liz89}).  One of the objectives of this paper is to 
develop a framework for determining
how these equilibria, particularly those at a
critical point, are affected by Alfv\'en waves.

\subsection{Pressures in Molecular Clouds}
\label{sec:pressures}

	GMCs appear to maintain a mean pressure that
is substantially above the pressure in the local interstellar medium.
Observations indicate that the mean kinetic pressures of nearby GMCs
range from $4\power4$ to $2.4\power5$
K~\cm\nthree\ (\cite{ber92}),
and the typical GMC in the molecular ring of the Galaxy has a mean
pressure of $3\power5$~K~\cm\nthree\ based on the column densities
inferred by Solomon et al.~(1987).
By comparison, the pressure in the local interstellar medium
is only about
$2\power{4}\ \K~\cm\nthree\ $ (\cite{bou90};
we have subtracted out the cosmic ray pressure since it is approximately
uniform and cannot contribute to the support of GMCs).

	The origin of the relatively high mean pressures in GMCs is
not understood.
Consider the simple, well--studied example of the isothermal sphere:
The maximum ratio of mean--to--surface 
pressure for a stable isothermal sphere
is only about 2.5
(e.g., \cite{spi68}),
well below what is deduced from observations.
The polytropes suggested by Maloney (1988)
have even smaller pressure ratios.
In the magnetized models of TIN, the maximum stable 
pressure ratio is also about 3.
Center--to--surface pressure ratios can be larger, but are
also often limited---for example,
a stable isothermal sphere has a maximum pressure drop 
of 14. 
As we shall see in
\S \ref{sec:locally}, however, it is possible to achieve
larger center--to--surface pressure drops (see also Curry \& McKee 1999).

	The resolution of this issue is crucial.
Because the mean pressure of GMCs is representative of the pressure 
of the star-forming regions within them, it is intimately coupled to the
star-formation process.
The mean pressure determines, in part,
the mean density in the star-forming region, which in turn affects
the density of the star cluster that forms from the cloud.
The line width--size relation (\cite{lar81}) suggests that that
the mean pressure of self-gravitating GMCs is constant:
In GMCs, the line width--size 
relation has the form $\sigma\propto R^{1/2}$, where
$\sigma$ is the 1-D velocity dispersion and 
$R$ the cloud radius ({\it e.g.}, \cite{sol87});
for self-gravitating clouds, we also have $\sigma^2\propto M/R$,
so that $M\propto R^2$, and hence
the mean pressure, ${\mathaccent22 P}\simeq{\mathaccent22
\rho} \sigma^2
\propto(M/R^3)R\propto R^0$, should be the same for all GMCs.
Chi\`eze (1987) and Elmegreen (1989) have suggested that this constancy
can be understood if the ambient pressure of GMCs is constant since,
in their models, ${\mathaccent22 P}$ is proportional to the ambient pressure.

	For both problems---the origin of the line width--size 
relation and the stellar number
density ultimately produced by the cloud---an understanding of the high
mean pressure of GMCs is essential.
Part of the explanation lies in the fact that the pressure
in the molecular gas is increased by the weight of the
overlying atomic gas.  Elmegreen (1989) pointed 
out the importance of this effect for the atomic hydrogen;
Holliman (1995) showed that the effect is even larger if
one is attempting to explain CO data on molecular clouds,
since the layer in which the carbon is atomic is
considerably thicker (and therefore heavier) than
the atomic hydrogen layer.  However, it is not
clear that this effect is sufficient.  Two other effects
that could contribute to the large pressures observed in GMCs are 
(1) the ambient pressures of GMCs could be several times greater than
the local interstellar value, 
perhaps due to the effects of massive star formation;
or (2) the structure of the clouds could permit
a mean pressure in the molecular gas
significantly above the values allowed in existing models.
The formalism developed herein can be used to evaluate the
effect of the overlying atomic gas and to determine the
mean pressure in the molecular gas.

\subsection{Previous Models of Self-Gravitating Gas Clouds}
\label{previous}

	In this paper we appeal to the relative simplicity of
polytropic spherical
models, in which $P(r)\propto\rho^\gp(r)$,
to determine the structure and stability of self-gravitating gas clouds.
Some of the earliest work on this problem was
carried out by Ebert (1955) and Bonnor (1956),
who investigated the stability of
pressure-bounded, gravitating spheres
supported exclusively by isothermal gas pressure.
These models become susceptible to dynamic instability when self-gravity
induces a center--to--surface density ratio exceeding $14$.
Non-isothermal polytropes ($\gp\neq 1$) have been considered 
by  Shu et al (1972),
Viala and Horedt (1974), and Chi\`eze (1987)
under the assumption that there is no internal heat flow
(a locally adiabatic system; see \S 2.2).  These results show that
the maximum stable center--to--surface pressure ratio increases with $\gp$.
Maloney (1988) considered polytropes with $\gp<1$ 
in order to account for the observed line width--size
size relation, since such polytropes have line widths that
increase outwards.  He assumed
that some unspecified heating mechanism was able to maintain a
constant central temperature when the cloud was compressed;
with this assumption, clouds are stable to gravitational collapse.   Lizano
\& Shu (1989), Gehman et al (1996), and McLaughlin \& Pudritz
(1996) considered polytropes in the limit $\gp\rightarrow 0$
(logatropes) to model turbulent pressure in clouds.
Isothermal polytropes ($\gp=1$) with non--isothermal
specific heats were analyzed by Yabushita (1968).

	Lynden-Bell and Wood (1968, hereafter LBW), 
used polytropes with $\gp=1$
to study self-gravitating star clouds (i.e., globular
clusters) in order to understand
the gravothermal catastrophe discovered by 
Antonov (1962). In contrast to the polytropic models of gas clouds cited
above, internal heat flow is allowed, although 
there is no heat transfer from the cloud to its
environment
(a globally adiabatic system; see \S 2.2).
Indeed, it is
efficient thermal conduction inside the star cluster
that establishes the isothermal temperature profile.
They modeled the cluster as a gas with no internal degrees of freedom.
If such a gas is confined by an ambient pressure,
the onset of dynamical instability occurs at a
center--to--surface density ratio of 389
(more precisely, we find 389.6);
the fact that the gas
heats up when it contracts permits the cloud to evolve to much greater
densities before it becomes unstable.
However, for a cluster of stars it is appropriate to
consider a 
gas confined to a fixed volume rather than
by a fixed pressure since
environment of the cluster
performs no work on the stellar fluid.
In this case, they showed that the maximum center--to--surface density
for a stable configuration is 709.

	The structure of rotating, self-gravitating clouds has been
discussed by Stahler (1983).
Most clouds in equilibrium are thought to derive significant support
from static magnetic fields, and magnetic braking should diminish the
dynamical influence of rotation (\cite{mes84}; \cite{mou87}).
The direct impact of static magnetic fields on the structure and stability
of gas clouds is explored in the axisymmetric models of Mouschovias (1976a,b),
Tomisaka \etal~(1988a), TIN, and Lizano \& Shu (1989).

	In \S3 we present an approach 
for determining the structure of a spherical
cloud supported by multiple pressure components.
The stability analysis in \S4 then allows us to determine the
range of stable equilibria that can represent cloud structure.
In many cases, these models are close to the critical
equilibrium that defines the limit of gravitational stability, since
it has been argued that 
ammonia cores (\cite{mye83}; \cite{fos93}), 
massive star--forming clumps (\cite{ber92}), and GMCs as a whole
(\cite{mck89}) all verge on gravitational collapse.

\section{Formulation of the Models} \label{s:set up}

\subsection{Multi--pressure Polytropes}

	Our intention in this work is to determine the	
the structure and stability of gas clouds 
in hydrostatic equilibrium that are supported by several 
pressure components.  The basic assumption underlying
our work is that such a quasi--static model can be
used to model turbulent molecular clouds.
We assume that in equilibrium,
each pressure component
satifies a polytropic relation of the form
$P_i(r)=\kpi\rho^\gpi(r)$, where $\kpi$ is independent
of position.  
In general, one can generalize the
classical analysis of polytropes to multiple components in
several ways:

\begin{enumerate}
\item{}{\it Multi--layered, or composite, 
polytropes} have spatially distinct pressure
components (Chandrasekhar 1939, p. 170), 
as in the core-envelope stellar models of 
Sch\"onberg \& Chandrasekhar (1942).  
Applications to gas clouds are considered
by Curry and McKee (1999).

\item{}{\it Multi-fluid polytropes} have several different components
that interact only gravitationally.  Examples include multi-mass
models for star clusters (Taff, Van Horn, Hansen, and Ross 1975),
models for clusters of galaxies including dark matter,
and models for molecular clouds including embedded stars.

\item{}{\it Multi--pressure polytropes} consist of a single 
self-gravitating fluid
with several pressure components, so that the total pressure
$P(r)$ satisfies the relation
\beq
P(r)=\sum P_i(r)=\sum \kpi\rho^\gpi(r).
\label{eq:pressure}
\eeq
\end{enumerate}
It is this last type of polytrope that is of interest
here.  As discussed above, the
pressure components 
relevant to molecular clouds are thermal pressure, static magnetic pressure,
and the pressure due to MHD waves.

	We shall restrict our attention to spherical polytropes.
Treating molecular clouds
as spherical is clearly a substantial idealization, but it
enables us to explore some of the essential physics underlying
their structure.
There are several effects that could lead to deviations from
spherical symmetry:  
(1) Rotation. As discussed in \S1.4 above, rotation is expected
to have a relatively small effect on the structure of molecular
clouds due to the effects of magnetic braking.
(2) Tidal gravitational fields. These also appear to
have a minor effect on Galactic molecular clouds
(Scoville \& Sanders 1987).
(3) Massive star formation.  Massive stars can significantly disrupt
molecular clouds, leading to violations of both hydrostatic
equilibrium and spherical symmetry.  Our models apply only to
clouds that are not being disrupted by this process;
fortunately, in the Galaxy there are many more GMCs than large OB
associations (Williams \& McKee 1997), so this will
be a reasonable approximation for many clouds.
(4) Anisotropic pressure.  The stress due to an ordered magnetic 
field is intrinsically anisotropic, and necessarily leads
to a violation of our assumption of spherical symmetry.
Nonetheless, many of the qualitative features of magnetic
stresses can be captured by treating them as spherically
symmetric (\cite{saf97}), and it is possible to
obtain reasonably accurate quantitative results for
the structure of magnetized clouds using the approximation
of a spherically symmetric magnetic stress (\cite{hol95}).
(5) Large scale turbulent motions can make clouds quite
non--spherical (\cite{bal99}); however, the time average
of the cloud shape is much closer to being spherical.

	Polytropes with a single pressure component
are often described
in terms of a parameter $n$ such that $\rho(r)\propto
T^n(r)$; the polytropic index $\gp$ is related to 
$n$ by 
\beq
\gp=1+\frac{1}{n}.
\eeq
``Negative--index polytropes"
with $n<0$ have been discussed by Shu et al (1972), Viala \& Horedt (1974),
and Maloney (1988).  Such polytropes are hotter on the outside
than at the center, just the opposite of a star.
An example of such a polytrope that is relevant for
our models is the pressure due to 
Alfv\'en waves, which satisfy $\gp=1/2$ (McKee \& Zweibel 1995), 
corresponding to $n=-2$.
We shall assume that $\gp>0$, since polytropes with $\gp<0$ are
unstable for all values of the external pressure (Viala \& Horedt
1974).

\subsection{Locally Adiabatic and Globally Adiabatic Pressure 
Components}

	In order to determine the stability
of multi--pressure polytropes, we 
assume that the response of each pressure component
to an adiabatic perturbation is the same as that of an
ideal gas that is either ``locally adiabatic"
or ``globally adiabatic."
Physically, a locally adiabatic pressure component has 
a time scale for internal
heat transfer that is long compared to the dynamical time scale,
whereas a globally adiabatic component has a short time scale
for internal heat transfer.
The distinction between local and global adiabaticity is necessitated
by the fact that we are treating self-gravitating systems that
are inhomogeneous.

	The response of a {\it locally adiabatic} pressure component to 
an adiabatic density perturbation satisfies the usual adiabatic relation,
\beq
\delta \ln P_i=\gi\delta\ln\rho.
\eeq
The specific entropy associated with a pressure component in our ideal
gas model is
\beq
s_i={k\over \gi-1}\;\ln\left({P_i\over\rho^\gi}\right) + {\rm const}
\label{eq:entropy}
\eeq
(e.g., Landau \& Lifschitz 1958).  The entropy is thus determined by
the {\it entropy parameter}
\beq
K_i\equiv {{P_i}\over{\rho^\gi}};
\label{eq:entparam}
\eeq
this does not change with time for a locally adiabatic pressure component.
Just as in the analysis of the stability of stars (\cite{led65}),
the adiabatic indexes $\gi$ that describe the temporal variations
of the pressure components need not
be the same as the polytropic indexes $\gpi$ that describe the
spatial variations.
The polytropic indexes $\gpi$ help to determine the equilibrium structure of a
cloud (\S3); stability depends additionally on the adiabatic 
indexes $\gi$ (\S 4).
Pressure components with $\gamma_i=\gpi$ are {\it isentropic},
since they have a spatially constant specific entropy that remains constant
during an adiabatic perturbation:
$K_i=P_i(r)\rho(r)^{-\gamma_i}\propto \rho(r)^{\gpi-\gi}$
is spatially and temporally constant for $\gamma_i=\gpi$.
Chandrasekhar (1939) termed such components as being in
{\it convective adiabatic equilibrium} since efficient
convection produces such a structure in a thermally insulated system.
On the other hand, if a locally adiabatic component has $\gamma_i\ne\gpi$,
then it cannot remain polytropic after a pressure perturbation;
Yabushita (1986) has considered this case for spatially isothermal
spheres ($\gp=1$).

	If internal heat transfer is significant for a pressure component,
then we assume that the component is {\it globally adiabatic}.
Such a pressure component retains its polytropic form during an
adiabatic perturbation due to efficient internal heat
flow in the cloud. 
The distinction between locally adiabatic and 
globally adiabatic components is important only for $\gi\neq\gpi$;
if the component is isentropic, then no heat transfer 
is required to maintain the polytropic structure, and 
the component is locally adiabatic as well.
The model of a globular cluster considered by LBW, in 
which the stars interact dynamically so as to maintain an isothermal
distribution, provides an example of a globally adiabatic system.
In our work, Alfv\'en waves, which
have $\gpi=1/2$ (\cite{wal44}, \cite{wei62}) 
and $\gi=3/2$ (\cite{mck95}), are modeled
as a globally adiabatic pressure component.
(Note that since $\gpi\neq 1$, the ``heat flow" associated with
the Alfv\'en waves does not make the gas isothermal.)
The polytropes considered by Maloney (1988), like our model of 
a polytrope supported by Alfv\'en waves,
have $\gpi<1$ and are not locally adiabatic;
however, Maloney's models require an injection of heat 
to maintain a constant central temperature during a compression, 
whereas our model
redistributes heat that is already present.

	It is important to distinguish the {\it physical} 
pressure components from the {\it model} 
pressure components.
We model each pressure component as a thermally insulated, ideal gas
with a local ratio of specific heats equal to $\gamma_i$.
If the pressure component is actually thermally insulated, then $\gamma_i$
is the ratio of specific heats, as usual.
On the other hand, if the gas is subject to heating and cooling,
then it can be modeled as a locally adiabatic component with
an adiabatic index $\gamma_i$ that is generally not
equal to the physical ratio of specific heats;
the effects of heating and cooling are treated
by invoking hypothetical internal degrees of freedom.
For example, if the rate for heating is proportional to $\propto nT^a$ and the
rate for cooling varies as $\propto n^2T^b$, then it can be shown that,
in equilibrium, the model has $\gamma_i=1+1/(a-b)$,
whereas the gas (if it is monatomic) 
actually has a ratio of specific heats of 5/3.
In such a case, $P_i(r,t)$ depends only on density, and
as a result, the pressure component is isentropic ($\gamma_i=\gpi$).

\section{Structure Equations} \label{s:structure}
 
	The elegance of polytropic models of self-gravitating clouds
is due to the small number of physical constants and
parameters that are needed to specify the 
structure of the cloud.
The structure is determined by the mass equation,
\beq
{dM\over d\R}=4\pi\R^2\rho,
\label{eq:dmdr}
\eeq
and the equation of hydrostatic equilibrium,
\beq
{dP\over d\R}=-{GM\over\R^2}\rho.
\label{eq:HSE}
\eeq
Note that these equations  remain valid 
even if the sphere evolves quasistatically.

	It is customary to combine these equations into a single second
order differential equation,
the Lane-Emden equation (e.g., \cite{cha39}; see 
Appendix \ref{app:a}).  However, 
this equation admits a homology transformation, so that
one of the two boundary conditions serves merely to set the
density scale (\cite{cha39}).  As a result, it is possible
to write this equation as a first order equation
in terms of the scale-free variables
\beq
u\equiv\frac{4\pi\rho r^3}{M(r)},~~~(n+1)v\equiv\frac{GM(r)\rho}{Pr},
\eeq
with a separate first order equation for the dimensionless radius
(\cite{cha39}).

	We have chosen a variation of this approach,
in which we place the two first order equations on an
equal footing.
We adopt 
\beq
\indep\equiv\ln(\rho_{\rm c}/\rho)
\label{eq:chi}
\eeq
as an independent variable,
where $\rho_{\rm c}$ is the central density of the {\it equilibrium}
cloud
(i.e., it does not vary when the cloud is perturbed).
$\chi$
is generally a monotonically increasing
function of $r$:
Since $P(r)$ is monotonically decreasing in a
self-gravitating cloud, $\rho(r)$ will also be monotonically decreasing
provided only that the pressure varies as a positive power of the
density ($\gpi>0$), which is necessary for stability in any case
(Viala and Horedt 1974).

	As dependent variables, we adopt one that
is proportional to the mass measured in Jeans masses
(cf Stahler 1983),
\beq
\mu\equiv {M(\R)\over c^3(r)/[G^{3}\rho(\R)]^{1/2}},
\label{eq:mu}
\eeq
and one proportional to the radius measured in Jeans lengths,
\beq
\lambda\equiv{\R\over c(\R)/[G\rho(\R)]^{1/2}},
\label{eq:lambda}
\eeq
where $c\equiv (P/\rho)^{1/2}$ is the isothermal sound speed.
More specifically, $\mu\simeq M(r)/M_{\rm J}(r)$,
where $M_{\rm J}(r)$ is the local generalized Jeans mass---i.e.,
the maximum mass that the concerted action of all
pressure components can maintain in equilibrium, given the values of
specified parameters, such as the entropy and ambient pressure.
Similarly, $\lambda\simeq r/R_{\rm J}(r)$, where
$R_{\rm J}(r)$ is the radius of a uniform-density sphere containing
$M_{\rm J}(r)$.
These variables are related to the standard homology variables
by
\beq
u=\frac{4\pi\lambda^3}{\mu},~~~(n+1)v=\frac{\mu}{\lambda}.
\eeq
An advantage of our dependent variables (and of the homology
variables) is that they depend upon the properties at the surface
of the cloud, not at the center.  For interstellar clouds,
the surface pressure is generally known, whereas the conditions at
the center of the cloud are not.
Further discussion of the 
relationship between our variables and the standard
\nondim\ variables can be found in Appendix
\ref{app:a}.

	Since individual polytropic pressure components 
can be expressed as $P_i=\kpi[\rho_{\rm c}{\rm exp}(-\indep)]^\gpi$,
the total pressure is 
\beq
P=\sum_i\kpi\rho_{\rm c}^\gpi e^{-\gpi\indep}.
\label{eq:P(chi)}
\eeq
Since $d\ln\rho=-d\indep$,
the overall polytropic index is
\beq
\gp\equiv{d\ln P\over d\ln\rho}=-\frac{d\ln P}{d\indep}=
	1-{d\ln c^2\over d\indep}.
\label{eq:gamma_p_I}
\eeq
For a single pressure component, $\gp$ is 
spatially constant for polytropes.  
However, for
a multi-component system, $\gp$ can be a function of position,
since equations (\ref{eq:P(chi)}) and (\ref{eq:gamma_p_I}) imply that
\beq
	\gp={\sum\gpi\kpi\rhoc^{\gpi}e^{-\gpi\indep}\over
	\sum\kpi\rhoc^{\gpi}e^{-\gpi\indep}}=
	\sum_i\pip \gpi~~~~~~({\rm polytrope}),
\label{eq:gpsum}
\eeq
where we have indicated that this equation
is valid only for polytropes, which have $\kpi$ constant.
In terms of our dimensionless variables,
the mass equation (\ref{eq:dmdr}) yields
\beq
{d\ln M\over d\indep}=4\pi\gp{\lambda^4\over\mu^2},
\label{eq:d_ln_M_2}
\eeq
whereas equation~(\ref{eq:HSE}) becomes
\beq
{d\ln\R\over d\indep}=\gp{\lambda\over\mu}.
\label{eq:d_ln_R_2}
\eeq
Appearances of $\ln M$ and $\ln\R$ in (\ref{eq:d_ln_M_2}) and
(\ref{eq:d_ln_R_2}) can now be eliminated in favor of expressions
involving $\mu$ and $\lambda$ with the aid of equations
(\ref{eq:mu}) and (\ref{eq:lambda}),
\beq
d\ln M=d\ln\mu+3d\ln c-{1\over2}d\ln\rho,
\label{eq:d_ln_M}
\eeq
and
\beq
d\ln\R=d\ln\lambda+d\ln c-{1\over2}d\ln\rho.
\label{eq:d_ln_R}
\eeq
We then use equations
(\ref{eq:gamma_p_I}), (\ref{eq:d_ln_M}), and (\ref{eq:d_ln_R})
in equations (\ref{eq:d_ln_M_2}) and (\ref{eq:d_ln_R_2})
to obtain the structure equations for multi--pressure polytropes:
\beq
{d\ln\mu\over d\indep}=4\pi\gp{\lambda^4\over\mu^2}
-\frac12(4-3\gp),
\label{eq:mu_diff}
\eeq
and
\beq
{d\ln\lambda\over d\indep}=\gp{\lambda\over\mu}
	-\frac12(2-\gp).
\label{eq:lambda_diff}
\eeq
If desired, one can obtain a single first order equation
by taking the ratio of these two equations;
once $\mu(\lambda)$, say, is found from integrating this equation,
the density can be evaluated by integrating equation
(\ref{eq:lambda_diff}).

	The structure equations are integrated outward from
the center of the cloud, which has $\mu=\lambda=0$ and $\indep=0$.
In order to interpret $\lambda/\mu$ near the center, we also need
a relation between these two variables there.  Since the cloud
is not singular, the density approaches a constant at the center,
and as a result
\beq
\frac{\mu}{\lambda^3}=\frac{M(r)}{\rho r^3}\rightarrow\frac{4\pi}{3}
	~~~~~~(\lambda\rightarrow 0).
\label{eq:muoverlam3}
\eeq

	Equations  (\ref{eq:mu_diff}) and
(\ref{eq:lambda_diff}), together with equation (\ref{eq:gpsum})
for $\gp$, describe the
the general structure of multi--pressure polytropes.
By inspecting these equations, we see that for a given $\gp$, 
equilibria are distinguished only by how far integration is carried in
$\indep$ and, thus, by how centrally concentrated they are.
Figure \ref{fig:ml} shows the structure of a spatially
isothermal sphere; the origin corresponds to the center of
the sphere, and the surface can be at any point on the curve.
The stability of the sphere is discussed in \S 4 below.

\subsection{Singular Polytropic Spheres}
\label{sec:singular}

	Just as the singular isothermal sphere
is very useful in studying the structure and dynamics of
isothermal clouds (Shu 1977), so the singular polytropic
sphere (which we label ``SPS") is useful in studying
the structure of polytropic clouds.  The basic relations
governing such spheres have been given by Chandrasekhar
(1939) and, more recently, by Stahler (1983) and
by McLaughlin and Pudritz (1996).
Singular spheres have power-law profiles for the density, etc.
The equation of hydrostatic
equilibrium (\ref{eq:HSE}) implies
\beq
\rho\propto r^{-2/(2-\gp)},~~~P\propto r^{-2\gp/(2-\gp)},~~~
	c\propto r^{(1-\gp)/(2-\gp)}.
\label{eq:plsps}
\eeq
Observe that the sound speed $c$ increases outward for
$\gp<1$, as discussed in \S 2.
In order for the mass to be finite, we require
$\rho$ to fall off more slowly than $1/r^3$, which implies $\gp<4/3$.
One can then readily show that an SPS is described by
single values of $\mu$ and $\lambda$
\beqa
\mu&=&\left(\frac{2}{\pi}\right)^{1/2}\frac{(4-3\gp)^{1/2}\gp^{3/2}}
	{(2-\gp)^2},
\label{eq:musps}\\
\lambda&=&\left(\frac{1}{2\pi}\right)^{1/2}\frac{(4-3\gp)^{1/2}\gp^{1/2}}
	{2-\gp}.
\label{eq:lamsps}
\eeqa

	An important parameter describing the thermal structure of
a cloud is the ratio of the mean square value of $c$,
\beq
\avg{c^2}\equiv\frac{1}{M}\int c^2 dM=\frac{1}{M}\int P dV
	=\frac{\bar P}{\bar\rho}
\label{eq:cavgt}
\eeq
to the surface value, where $\bar P$ and $\bar\rho$ are
the volume-averaged pressure and density, respectively.
For an SPS, we find
\beq
\psi\equiv\frac{\avg{c^2}}{c_\s^2}=\frac{4-3\gp}{6-5\gp},
	~~~~~~~~(\gp<\frac{6}{5}),
\label{eq:psi}
\eeq
where we have added the subscript ``s" on $c^2$ to emphasize
that it is measured at the surface of the cloud.

     Note that $\psi$, which is equivalent to the mean energy
per gram divided by the surface value,
diverges as $\gp\rightarrow 6/5$: 
for $\gp\geq 6/5$, the energy of an SPS is concentrated
at the center.  Physical polytropes must satisfy the boundary
condition $dP/dr=0$ at the origin, since the gravitational force
vanishes there.  The fact that SPSs do not satisfy this
condition is not important for $\gp<6/5$, since the mass and energy 
of such polytropes are dominated by the outer layers.
As a result, the solutions for physical polytropes approach
the SPS solutions for large values of $\chi$.
On the other hand, for $\gp\geq 6/5$, the fact that 
SPSs do not satisfy the proper central boundary condition means that
they cannot serve as approximations for physical polytropes.
For example, equation (\ref{eq:plsps}) shows that an SPS
with $\gp=6/5$ has $\rho\propto r^{-5/2}$, whereas
the analytic solution for this case gives $\rho\propto
r^{-5}$ at large radii (Chandrasekhar 1939).

   The mean density in an SPS is
\beq
\frac{\bar\rho}{\rho_\s}=\frac{3\mu}{4\pi\lambda^3}
	\rightarrow \frac{3(2-\gp)}{(4-3\gp)}~~~~~~~~(\gp<\frac 43),
\label{eq:rhobar}
\eeq
whereas the mean pressure is
\beq
\frac{\bar P}{P_\s}=\frac{3\mu\psi}{4\pi\lambda^3}
	   \rightarrow \frac{3(2-\gp)}{(6-5\gp)}~~~~~~~~(\gp<\frac 65).
\label{eq:pbar}
\eeq
(In each case, the first equation is general and the expression
after $\rightarrow$ is for the SPS.)  As emphasized in
the Introduction, polytropes do not have large mean
density or pressure contrasts with the ambient medium:
for $\gp<1$, both are less than a factor 3 for an SPS.

   The gravitational energy $W$ can be described by the parameter $a$
defined by
\beq
W\equiv -\frac{3}{5}a\frac{GM^2}{r}.
\eeq
For an SPS, we have
\beq
a=\frac{5}{3}\left(\frac{4-3\gp}{6-5\gp}\right)=\frac{5}{3}\;\psi,
	~~~~~~~~(\gp<\frac 65).
\eeq

\subsection{The Virial Theorem}
\label{sec:virial}

	Before passing on to a consideration of the stability of
multi--pressure polytropes, it is worthwhile to determine
the implications of the virial theorem for our problem.
The virial theorem for an unmagnetized gas sphere is
\beq
3\int PdV-3PV+W=0.
\label{eq:virial}
\eeq
Note that this equation applies at any point within
the sphere as well as at its surface.
In our notation, this becomes
\beq
3\psi\lambda\mu-4\pi\lambda^4-\frac{3}{5}\, a\mu^2=0.
\eeq
This can be solved to give an explicit relation for
$\mu(\lambda)$ in terms of two parameters, $a$ and 
$\psi$, that are usually of order unity:
\beq
\mu=\frac{5\psi\lambda}{2a}\left[1\pm\left(1-\frac{\dis 16
	\pi a\lambda^2}{\dis 15\psi^2}\right)^{1/2}\right].
\label{eq:muoflambda}
\eeq
Note that $\mu$ has the correct limit (\ref{eq:muoverlam3})
as $\lambda\rightarrow 0$ if the minus sign is chosen
(since $a$ and $\psi\rightarrow 1$ as $\lambda\rightarrow 0$).
The solution with the minus sign 
is the correct solution for all values of $\lambda$ for
$\gp<0.82$; it is correct for all stable
isentropic spheres for $\gp<0.90$.
For $\gp>0.82$, the solution can
switch sign as $\lambda$ increases:  
For $0.87>\gp>0.82$ there
can be multiple sign changes, whereas for $\gp>0.87$ there is
only one change in sign, from minus to plus.
Equation (\ref{eq:muoflambda})
is exact, and it is not restricted to polytropes.

\section{The Stability of Multi--pressure Polytropes} 
\label{sec:stability}

\subsection{General analysis}
\label{sec:general}

	LBW have given a general discussion of the 
gravitational stability of 
spatially isothermal spheres ($\gp=1$).
They considered both the case of a sphere subject to a 
given pressure, which is appropriate for a cloud of gas, 
and that of a sphere confined to a 
fixed volume, which is a model for a cluster of stars.  
They focused on the cases $\gp=\gamma=1$ (an isothermal sphere,
which is the Bonnor--Ebert problem) and $\gp=1$, $\gamma=5/3$.
Here, we shall generalize their discussion of pressure-bounded spheres
to the case of arbitrary $\gamma$ and $\gp$.
Our objective is to find the criterion for determining the 
{\it critical point}, which is the most centrally concentrated 
structure that is gravitationally stable.
Equivalently, the critical point gives the maximum mass for which an
equilibrium exists at the specified conditions.

	The first law of thermodynamics for the cloud reads
$\delta E=\delta Q-P_\s\delta V$, where $E$ is the total energy
of the cloud, $\delta Q$ is the net heat flow
into the cloud, and $P_\s$ is the pressure at the surface of the cloud.
We shall focus on systems that are in equilibrium.
Such systems are characterized by having a 
stationary value of the
entropy subject to the given constraints. 
The equilibrium is stable if the entropy is a maximum.

	We wish to determine the stability criterion for a 
pressure-bounded gas cloud.
The stability of a system is determined by its thermodynamic free energy.
For an adiabatic, pressure-bounded system, this free energy is the enthalpy,
$H=E+P_\s V$, the sum of the energy of the cloud and the work required
to displace the volume occupied by the cloud.
Note that $P_\s$ is the pressure at the surface of the cloud, so that
(in contrast to the energy) the enthalpy of the cloud is not
calculated as an integral over the elements of gas within the cloud.
Furthermore, as discussed in \S2.2, an appropriate choice of the
adiabatic index $\gamma_i$ often 
allows one to model pressure components
that are subject to heating and cooling as being adiabatic.
Using the first law of thermodynamics, we find for the cloud 
\beq
\delta H=\delta Q+V\delta P_\s.
\label{eq:delta enthalpy}
\eeq

	Following LBW, we assess stability by considering a linear 
series of equilibria,
an approach originally developed by Poincar\'e (1885) for mechanical systems.  
Consider a series of gas spheres, all of which are in equilibrium and all 
of which represent clouds that have the same 
mass and entropy, but that are confined
by an ambient pressure $P_\s$ that varies monotonically along the series.  
(For example,
this series of equilibria could result from the adiabatic compression
or decompression of a cloud.)
For spheres supported by more than one pressure component, we assume
that the entropy of each component is constant along the series.
These gas spheres need not be polytropes; this analysis thus
applies to perturbed locally adiabatic spheres,
which are not polytropes if $\gamma\neq\gp$.
In some cases, one will reach a critical point beyond which no equilibrium
exists; i.e., $\delta P_\s=0$ at the critical point.
This point is an extremum and not a saddle point 
by construction, since we have assumed
that there are no equilibria beyond the critical point;
it will be a maximum or a minimum  depending on whether the pressure must
be increased or decreased to reach the critical point.
Since the condition $\delta P_\s=0$ cannot
distinguish between an extremum and a saddle point, 
it is a necessary, but not
sufficient, condition for a critical point.
Now recall that in the series of equilibria that we have been considering
no heat has entered or left the system ($\delta Q=0$).
It therefore follows from equation (\ref{eq:delta enthalpy}) that 
we also have $\delta H=0$ at the critical point.

	Alternatively, one can fix the pressure and gradually change 
the entropy
by allowing heat to flow into or out of the cloud.
Again, depending on the values of $\gamma$ and $\gp$, one may reach a
critical point beyond which no equilibrium exists.
In this case $\delta Q=0$ when one reaches the critical point; 
and since we assumed $\delta P_\s=0$, it follows that $\delta H=0$
at the critical point
for this sequence of equilibria as well.
For simplicity, however, we shall focus on adiabatic pressure perturbations.
Even though we assume that the perturbations in our model
are adiabatic, we can treat heat transfer from the environment by
allowing $\g$ to differ from the physical ratio of specific
heats (see \S 2.2).

	A complementary approach is to consider a sequence of spheres
with a monotonically varying mass,
all with the same entropy (or entropy per unit mass) and all
embedded in a medium of constant pressure.  In this case,
the critical point corresponds to the sphere with an extremum in
the mass ($\delta M=0$).  For self-gravitating clouds, this
extremum (if it exists) is the maximum stable mass under the
specified conditions, and we label this $\Mcr$.
Since $\delta M=0$ at the critical point,
it does not matter whether it is the entropy or entropy per unit
mass that is held constant along the sequence.\footnote{The 
approach of considering a sequence of equilibria
is also used in assessing the stability of degenerate stars
(\cite{sha83}).
Stars have zero pressure at their surfaces,
so the criterion $\delta P_s=0$ that we are using for clouds
is irrelevant.  Although degenerate
stars have negligible entropy, they have a well-defined entropy
parameter $K$ and adiabatic index $\gamma$, 
which in general depend on the density
but approach constants in the non-relativistic and
extreme relativistic limits.  The mass sequence can 
be arranged with either the central density or the radius
as the independent parameter.  The dependence of the equation
of state on density leads to the possibility of multiple
critical masses along the sequence (the Chandrasekhar mass
and the maximum neutron star mass).  In principle, 
such behavior can occur for multi--pressure polytropes as well.}

	The existence of a critical point depends on the 
value of $\gamma$ and on the structure of the cloud,
which is parameterized by $\gp$ for a polytrope.
For example, isothermal spheres ($\gamma=\gp=1$; see Fig. \ref{fig:pv}), 
in which adiabatic compression increases the magnitude of the
gravitational energy faster
than the thermal energy, have a critical point, whereas spheres with
$\gamma=\gp>4/3$, in which the opposite is true, do not.
Bonnor (1956) and Ebert (1955) used the condition $\delta P_\s=0$
to determine the stability of isothermal spheres.

	  Our approach in determining the stability
of self-gravitating clouds is quite different from
that of Maloney (1988) or McLaughlin and Pudritz
(1996).  
Because these authors used a phenomenological equation of
state, they could not define the entropy.
As a result they were forced make an arbitrary assumption
in specifying the applied perturbation;
they chose to keep the central temperature
constant.  While this assumption may be plausible for the thermal pressure
of cold molecular gas with $T\simeq 10$ K, it is unjustified
for wave pressure.  Indeed, if
the perturbed cloud is to remain polytropic for $\gp<1$, as they assumed,
then $\gamma=\gp$ and the cloud would cool as it is compressed;
in order to maintain a constant central temperature, 
heat must be supplied throughout the cloud 
to counteract this adiabatic cooling.
No mechanism has been advanced to account for such behavior, and
we regard it as unphysical.

\subsubsection{Isothermal Spheres}

	The special case of an isothermal sphere 
is frequently considered in the interstellar literature.
Such a sphere has a spatially constant temperature ($\gp=1$)
and its temperature remains constant if it is perturbed ($\gamma=1$).
By setting the ratio of specific heats for the gas to unity,
we can treat isothermal pressure perturbations as being 
adiabatic; the work done
on the gas goes into the hypothetical internal degrees of freedom.
As a result,
the critical point for this problem can be determined by extremizing $H$
in the usual manner.
It is customary, however, to use the Gibbs free energy $G=H-TS$ and
an adiabatic index equal to the actual ratio of specific heats.
For an isothermal sphere, $\delta Q=T\delta S$, so that
equation (\ref{eq:delta enthalpy}) implies $\delta 
G=-S\delta T +V\delta P_\s$.
When $\delta T=0$, the critical point, which is characterized
by $\delta P_\s=0$, also corresponds to the condition $\delta G=0$.
Either way, one finds that the 
maximum possible density 
contrast for a stable isothermal sphere is 14.0 (Bonnor 1956, Ebert 1955).

\subsubsection{Global Collapse vs. Core Collapse}
\label{sss:Global Collapse vs. Core Collapse}

	When an isothermal sphere collapses, the collapse is global:
all flow is inward.
On the other hand,
when a globular cluster collapses, the collapse is restricted to
the core; the envelope actually expands due to the transfer of heat
from the core (LBW).
Such a heat transfer is generally required to effect 
core collapse, and as a result it cannot occur in
clouds with only locally adiabatic pressure.
For globally adiabatic clouds, the 
dividing line between core collapse and global
collapse occurs at $\g=4/3$.
Globally adiabatic spheres with $\gamma=4/3$ evolve homologously
(\cite{sha83}):
If one were to compress a $\g=4/3$ sphere, the rate at which
the work done on the sphere is transformed into kinetic energy
would exactly balance the rate of increase of the magnitude of 
the gravitational energy.
Because the fraction of
$PdV$ work that is converted to kinetic energy
increases with $\gamma$, compression of a sphere with
$\g>4/3$ everywhere increases the ratio of kinetic to gravitational
energy, 
reducing the center--to--surface density ratio $\rhoc/\rhos$.
Conversely, if the surface of such a 
globally adiabatic
sphere were to expand,
the ratio of kinetic to gravitational energy would decrease.
If this evolution were carried sufficiently far,
the cloud would become unstable, resulting in collapse of the core
and dynamical expansion of the envelope.  Thus, for
$\gamma>4/3$, gravitational collapse is due to
rarefaction of the cloud, not compression.

	Globally adiabatic clouds 
with $\g<4/3$ everywhere exhibit the opposite
behavior.
For such clouds, compression increases the magnitude of the
gravitational energy faster than it does the kinetic energy,
so that $\rhoc/\rhos$ increases.
If the compression is
sufficiently great, the cloud becomes unstable, and the
entire cloud collapses.

\subsection{Equations for the Perturbed Structure}
\label{sec:equations}

	In order to assess the stability of multi--pressure
polytropes quantitatively, it is necessary to perturb
the structure equations.  This is done in Appendix
\ref{app:derivation}; we summarize the results here.
For simplicity, we assume that there is only one
globally adiabatic component.

	 The equations governing the variations in $\mu$ and
$\lambda$ are
\beq
{d\delta\ln\mu\over d\indep}=2\delta\ln\gp+(4-3\gp)(2\dll-\dlm)
\label{eq:ddlm},
\eeq
and
\beq
{d\delta\ln\lambda\over d\indep}=
	\delta\ln\gp+\left[\gp\left(\frac{\lambda}{\mu}+1\right)-2\right]
	\dlm+\left[4-3\gp\left(\frac{\lambda}{\mu}+\frac23\right)
	\right]\dll.
\label{eq:ddll}
\eeq

	These equations involve the perturbation in
the polytropic index, $\delta\gp$, which is also
evaluated in Appendix \ref{app:derivation}.  There
we find it convenient to introduce a new adiabatic index
\beq
\Gamma\equiv\sum_\ell \pip\gamma_i + \pgp\gpg,
\eeq
which is the weighted mean of the adiabatic indexes for the locally
adiabatic components $\ell$ and the polytropic index for the globally
adiabatic component $g$.  In terms of $\Gamma$, one finds
\beq
\delta\ln P=\frac{2}{4-3\Gamma}\left[\Gamma\dlm
	  +2\pgp\dlkg\right].
\label{eq:dlp}
\eeq
This result together with equation
(\ref{eq:deltaxi2}) enable us to evaluate $\gamma$,
the adiabatic index for a multi--pressure polytrope.
In the simple case in which there is no globally adiabatic component,
we find
\beq
\gamma\equiv\frac{\delta\ln P}{\delta\ln\rho} =
	-\frac{\delta\ln P}{\delta\chi}
	=\Gamma=\sum_\ell\pip\gamma_i~~~~~~~(P_g=0),
\eeq
which is just the weighted mean of the individual values of $\gamma_i$;
thus, in this case, $\Gamma$ reduces to $\gamma$.  On the other hand,
if there is a globally adiabatic pressure component but
the locally adiabatic components are isentropic, then
$\Gamma=\gp$.

     In Appendix \ref{app:derivation} we also define
\beq
\Gamma'\equiv\sum_{i\in l}\pip\gamma_i(\gp-\gpi) + 
	\pgp\gpg(\gp-\gpg).
\eeq
In terms of $\Gamma$ and $\Gamma'$, we find
\beq
\delta\ln\gp=\frac{1}{\Gamma}\left[4(\gp-\Gamma)\dll
	-2\left(\gp-\Gamma+\frac{\Gamma'}{4-3\Gamma}\right)\dlm
	+\left(\gpg-\gp-\frac{3\Gamma'}{4-3\Gamma}\right)\pgp\dlkg\right],
\label{eq:delta_ln_gp}
\eeq
where $\dlkg$ is the perturbation in the polytropic
coefficient for the globally adiabatic component and is
evaluated in Appendix \ref{app:global}.
If there are no globally adiabatic components, then 
this expression simplifies to
\beq
\delta\ln\gp=\frac{1}{\gamma}\left[ 4(\gp-\gamma)\delta\ln\lambda
	-2\left(\gp-\gamma+\frac{\Gamma'}{4-3\gamma}\right)
	\delta\ln\mu\right].
\eeq

	The solution for the perturbed structure is then given
by inserting equation (\ref{eq:dlk2}) into equation
(\ref{eq:delta_ln_gp}), then inserting that into equations 
(\ref{eq:ddlm}) and (\ref{eq:ddll}) for $\delta\ln\mu$ and $\delta\ln\lambda$,
and then solving them simultaneously with equations (\ref{eq:mu_diff}) 
and (\ref{eq:lambda_diff}) for $\mu$
and $\lambda$.  This procedure is more complicated than
the conventional treatment of the stability of hydrostatic clouds,
which permits the stability to be determined from the solution
of a single second order equation involving the spatial gradient
of the adiabatic index $\gamma$ (Ledoux 1965). The additional
complexity arises both because we have dropped the assumption that the
perturbations are locally adiabatic, and because we have
in effect evaluated the gradient of $\gamma$.

	As we found in \S 4.1, the critical point
is given by the condition that 
$\delta\ln P_\s=0$.  Equation (\ref{eq:dlp}) then implies that
the critical point is determined by 
\beq
\dlm=-\frac{2}{\Gamma}\pgp\dlkg.
\label{eq:critpt}
\eeq
This condition can be used to determine the critical
mass for clouds supported by locally adiabatic pressure
components,
for those supported by globally adiabatic pressure components,
and for multi--pressure polytropes that may be supported by
both types of pressure.

\section{Locally Adiabatic Polytropes}
\label{sec:locally}

	In multi--pressure polytrope models of GMCs, the
thermal gas pressure and the magnetic stresses are
modeled as locally adiabatic pressure components:
The entropy associated with the component remains
constant in each mass element during a perturbation.
Here we first determine how to characterize
the average entropy associated with a 
locally adiabatic pressure component, and then
use this to describe the critical mass.  As we
shall see, non--isentropic clouds are particularly
interesting since they can have large central--to--surface 
pressure ratios. 

\subsection{Entropy of Locally Adiabatic Pressure Components}
\label{sec:entropy1}

	For locally adiabatic
pressure components (labelled by $\ell$), 
the entropy parameter
$K_\ell=P_\ell/\rho^\gl$ remains constant in each mass element
during the evolution of the cloud (see eq. \ref{eq:entropy}
for the relation between $K_\ell$ and $s_\ell$,
the specific entropy associated with pressure component $\ell$).  
In general, $K_\ell$
depends on position unless the system is isentropic ($\gl=\gpl$).

	Particularly for the case in which magnetic
fields are important, it is
convenient to introduce a 
spherically symmetric reference state
of constant pressure $\plref$ (Mouschovias 1976a). 
We shall generalize Mouschovias' concept of
a reference state by allowing the density
in the reference state $\rhoref$ to depend on position in
the cloud; we do not assume that the reference state
is a polytrope.
For a locally adiabatic component,
the entropy parameter is the same for each mass element
in the cloud as it is for that mass element in the 
reference state, 
\beq
K_{\ell}(M)\equiv\frac{P_\ell(M)}{\rho(M)^\gl}=\frac{P_{\ell, \rm
	ref}}{\rhoref(M)^\gl}.
\label{eq:kl}
\eeq
In the isentropic case ($\gl=\gpl$),
the reference density is constant,
as Mouschovias assumed:
since the entropy
parameter $K_\ell$ is identical to the factor
$\kpl$ that governs the polytropic structure,
it is constant; with both $K_\ell$ and $\plref$ constant,
it follows that $\rhoref$ is as well.

	For non-isentropic systems,
we define the average entropy parameter by
\beq
\kavl^{1/\gl}\equiv\frac{1}{M}\int K_\ell^{1/\gl} dM.
\label{eq:kav1}
\eeq
Introducing the power $1/\gl$ into the definition
enables us to evaluate the integral for the reference state,
\beq
\kavl=\frac{P_{\ell, \rm ref}\vref^\gl}{M^\gl}=
	\frac{P_{\ell, \rm ref}}{\bar\rhoref^\gl},
\label{eq:kav2}
\eeq
where $dM=\rhoref d\vref$ and $\vref$ is the volume in
the reference state.  Thus $K_\ell^{1/\gl}$ is a measure of the
local entropy per unit mass, and $\kavl^{1/\gl}$
is a measure of the mean entropy per unit mass of an entire
cloud.

	For an isothermal gas, we have $\pth=\rho\cth^2$
with $\cth=$const, corresponding to $\gl=1$.  As a result,
the entropy parameter is simply 
\beq
K_{\rm th}=\avg{K_{\rm th}}=\cth^2.
\label{eq:kth}
\eeq

	Next, consider a cloud supported by a poloidal magnetic field.
We treat the magnetic
field as a gas with an adiabatic index $\gamma_B=4/3$.
This relation is exact for a tangled field.  For a poloidal
field, it is approximate, but it ensures that
the magnetic flux $\Phi=\int 2\pi rB dr$ is conserved during a 
homologous, spherical
compression, since then $ B\propto \rho^{2/3}$ and
$ rBdr \propto r^{2}\rho^{2/3}(dr/r)$=const.  
The parameter that measures the mean specific entropy 
for the field is 
\beq
\kavb=\frac{\pbref V_{\rm ref}^{4/3}}{M^{4/3}}
	\propto\frac{B_{\rm ref}^2R_{\rm ref}^4}{M^{4/3}}
\eeq
in terms of quantities measured in the reference state.
Since the flux $\Phi$ is the same in the cloud and
in the reference state, we can rewrite this as
\beq
\kavb\propto \frac{\Phi^2}{M^{4/3}}\propto\frac{\bar B^2}{\bar\rho
	^{4/3}},
\eeq
where $\bar B\equiv \Phi/\pi R^2$ is the mean magnetic field
and $\bar\rho$ is the mean density.
It follows that the parameter that measures the total entropy of
the cloud depends only on the magnetic flux, 
\beq
M\kavb^{3/4}\propto \Phi^{3/2}.
\eeq
Adiabatic evolution of the cloud thus corresponds to evolution 
with constant magnetic flux.

\subsection{The Critical Mass}
\label{sec:critical1}

	As discussed in \S4.1, the
critical mass is the mass of a cloud at the critical point:
For given values of the ambient pressure and component
entropies, there is no nearby equilibrium state with a 
mass greater than $\Mcr$.  
In order to determine the value of the critical
mass, we use the approach adopted in \S 4, based on
a sequence of equilibria with constant
mass and entropy but variable pressure.
(It should be kept in mind that this sequence of equilibria
is not a sequence of polytropes for non-isentropic clouds.)
For simplicity, we consider a
cloud supported by a single pressure component.
We can express the mass of the cloud (eq. \ref{eq:mu}) 
in terms of the entropy parameter
at the cloud surface, $K_{\ell\s}$, as
\beq
M=\frac{\mu}{G^{3/2}}P_\s^{(3\gl-4)/(2\gl)}K_{\ell\s}^{2/\gl},
\label{eq:Mkls}
\eeq
where we have introduced the subscript ``s" to emphasize
that the quantities are evaluated at the surface of the cloud.
The critical point occurs at an extremum of the surface
pressure, $\delta P_\s=0$.  If we adiabatically perturb a cloud
of fixed mass, equation (\ref{eq:Mkls}) shows that the critical point
occurs at
\beq
\dlm=0.
\label{eq:LAcritpt}
\eeq
This is just the condition implied by equation
(\ref{eq:critpt}) in the absence of
globally adiabatic components.
The critical mass is then
\beq
\Mcr=\frac{\mucr c_s^{3/2}}{G^{3/2}\rho_\s^{1/2}}
		 =\frac{\mucr c_s^2}{G^{3/2}P_\s^{1/2}},
\label{eq:Mcr}
\eeq
where $\mucr$ is a number that depends on $\gl$ and $\gpl$.

	For a non-isothermal gas, it is convenient
to have an expression for the critical mass in terms of
$\cavt\equiv M^{-1}\int c^2 dM$, 
which is often more readily observable.  To accomplish this,
we re-express equation (\ref{eq:mu}) as
\beq
M=\frac{\mu'\cavt^{3/2}}{G^{3/2}\rho_\s^{1/2}},
\label{eq:Mmup}
\eeq
where
\beq
\mu'\equiv\frac{\mu}{\psi^{3/2}};
\label{eq:mup}
\eeq
recall that $\psi\equiv\cavt/c_\s^2$ (see eq. \ref{eq:psi}).
At the critical point, we have $\mucr'=\mucr/\psi_{\rm cr}^{3/2}$.
For an isothermal cloud, $\psi=1$ and $\mucr'=\mucr$.

	It is important to keep in mind that $\Mcr$ is the maximum
stable mass for given values of $\avg{K_\ell}$ (or, equivalently,
$K_{\ell\s}$) and $P_\s$; it is the maximum stable mass
for given values of
of $c_\s$ and $P_\s$ (or, equivalently, $\rho_\s$)
only for isentropic spheres.  In \S 5.3.1 below
we show that $\mu$ is indeed a maximum at the critical point for
isentropic spheres.  For non--isentropic
spheres, which are discussed in greater detail
below (\S \ref{sec:non}), $\mucr$ is less
than the value for isentropic spheres.
This can be seen graphically in Figure \ref{fig:ml},
which portrays all the equilibria for isothermal
spheres:  As $\gl$ rises above unity, our numerical
results show that the critical
point moves down and to the left, following this
curve as the maximum stable value of $\chi$ increases.
(This result for isothermal spheres was first found
by \cite{yab68}.)  How is it then possible to 
satisfy the critical point condition $\dlm=0$
at a point other than the maximum in Figure \ref{fig:ml}?
The answer is that the perturbed sphere is not
a polytrope for $\gl\neq \gamma$ and is therefore
not represented in this figure.

	The expressions for the critical mass above depend
on the {\it specific} entropy.  It is also possible
to express the critical mass in terms
of the {\it total} entropy in the cloud, which is related
to $M\kavl^{1/\gl}$.  We define this alternative form for
the critical mass, which we label $\Mcrt$, by
\beq
\left(\frac{\Mcrt}{M}\right)^3\equiv\frac{\Mcr}{M}.
\label{eq:Mcrt}
\eeq
One can show that
$\Mcrt$ indeed depends on $M$ and $\kavl$
only through the combination
$M\kavl^{1/\gl}$ with the aid of equation (\ref{eq:Mkls});
it also depends on the numerical parameter $(K_{\ell\s}/\kavl)_{\rm cr}$.
At the critical mass ($M=\Mcr$), the two forms
for the critical mass are identical ($\Mcr=\Mcrt$).

	For an isothermal gas, the critical mass is the
Bonnor--Ebert mass,
\beq
\Mcr=\mbe=1.1822\left(\frac{\cth^3}{G^{3/2}\rho_\s^{1/2}}\right),
\eeq
which follows from equation (\ref{eq:Mcr}) with
$\mu_{\rm cr}=1.1822$.  
The alternate form for the critical mass,
$\Mcrt$, depends on the total thermal energy
in the cloud, $M\cth^2$.  

	Magnetically supported clouds have $\gamma=4/3$ as
discussed above.
The critical mass can thus be expressed as
\beq
M_B\equiv \Mcr\propto \left(\frac{K_{B}}{\kavb}\right)^{3/2}_{\rm cr}
	\frac{\kavb^{3/2}}{G^{3/2}}
	\propto\frac{\bar B^3}{G^{3/2}\bar\rho^2},
\eeq
where we have dropped the numerical factor $(\dots)_{\rm cr}$ in 
the last step.  Note that because $\gamma=4/3$, the
critical mass (eq. \ref{eq:Mkls} evaluated at the critical point)
is {\it independent of
the pressure}: for $M>M_B$, clouds are magnetically supercritical
and cannot be prevented from collapsing by the magnetic field,
whereas the contrary is true for $M<M_B$, under the assumption of
flux freezing (cf. Shu et al 1987).
The alternative form for the critical mass in terms
of the parameter that measures the total entropy is
\beq
M_\Phi\equiv\Mcrt =(M^2\Mcr)^{1/3}
	\propto\left(\frac{M^2\kavb^{3/2}}{G^{3/2}}\right)^{1/3}
	\propto\frac{\Phi}{G^{1/2}}.
\eeq
The numerical value for $M_\Phi$ depends on the distribution
of the flux in the cloud; for the standard case in which
a uniform field threads a spherical cloud, it is given as
$0.126\Phi/G^{1/2}$ by Mouschovias \& Spitzer (1976). 
The relation between the two forms of the magnetic critical
mass appears to have been first given by Mouschovias and
Spitzer (1976); equation (\ref{eq:Mcrt}) shows that
this relation is a general one that applies to all
forms of support by an adiabatic pressure.

\subsection{Results for Single Component, Isentropic Polytropes
($\g=\gp$)}
\label{sec:results1}

	The pressure components in interstellar clouds
are ``soft", with polytropic indexes $\gpi\leq 4/3$,
and this constrains the mass, density and pressure
of self--gravitating interstellar clouds;
for example, a stable isothermal sphere must be
less than the Jeans mass and has a maximum density
ratio of 14.04.  Here we consider the limitations
on the mass of single component, isentropic polytropic
polytropes with $\gp<4/3$.

	The stability analysis of isentropic polytropes,
previously considered by Shu et al (1987), Viala \& Horedt (1974) and  
Chi\`eze (1987), is a subset of our work, and we have plotted the
results in Figures {\ref{fig:rho} and {\ref{fig:rhobar}.
Note that
the critical density and pressure ratios increase with $\gamma$.
The points marked on these figures
denote the values for an isothermal sphere,
the case originally considered by Bonnor (1956) and Ebert (1955).
When $\gamma>4/3$, an isentropic polytrope is unconditionally stable.
Such a polytrope can be stable even if the density and pressure
vanish at its surface, and as such they can provide a simple
model for stars.
Systems with $\g=4/3$ evolve homologously in response to either a pressure
perturbation or an entropy perturbation (\cite{sha83}).
The critical configuration for such a
polytrope has a vanishing density at the surface and
is neutrally stable against collapse.
For example, white dwarf stars supported by ultrarelativistic degeneracy
pressure ($\g=4/3$)
are neutrally stable, their stability being dependent on mass but not
radius.
Polytropes with $\g=4/3$ and with a finite density at the surface
are unconditionally stable against collapse.

	The critical point condition $\delta\ln\mu=0$
(eq. \ref{eq:LAcritpt}) enables us to derive a simple
relation between $\lambda$ and $\mu$ at the critical point for
single component isentropic polytropes.  Such
clouds have constant $\gp$, so integration of
equations (\ref{eq:mu_diff}) and (\ref{eq:lambda_diff}) yields a
unique solution.  The perturbed cloud is also isentropic, so the
perturbed cloud must lie on this solution as well.  As a result,
isentropic clouds have identical
spatial and Lagrangian variations, so that $\delta\ln\mu=d\ln\mu=0$ at the
critical point.  (Recall from the discussion in \S \ref{sec:general}
that this extremum is a maximum.)
Equation (\ref{eq:mu_diff}) then implies
\beq
{\mu_{\rm cr}^2\over\lambda_{\rm cr}^4}={8\pi\gp\over4-3\gp}
\mbox{~~~~~~($\gp=\gamma$)}.
\label{eq:critical relationship}
\eeq
Chi\`eze (1987) previously obtained this relation 
(his equation A18)
when he considered the stability of 
single component polytropes, under the implicit
assumption that $\gamma=\gp$.
The equivalence of the equations follows from the variable
transformations in Appendix \ref{app:a}.
With the aid of the expression for $\mu$ obtained
from the virial theorem (eq. \ref{eq:muoflambda}) it
is possible to evaluate $\mu_{\rm cr}$ explicitly,
\beq
\mu_{\rm cr}=\frac{9}{8(2\pi)^{1/2}}\;\frac{\psi_{\rm cr}^2
	(4-3\gp)^{1/2}\gp^{3/2}}{\left[1-\frac 34\left(1-
	\frac 25\, a_{\rm cr}\right)\gp\right]^2}
	\mbox{~~~~~~($\gp=\gamma$)}.
\label{eq:mucr}
\eeq
The parameters $a_{\rm cr}$ and $\psi_{\rm cr}$
are of order unity for $\gp<1.3$; as $\gp\rightarrow 4/3$,
we find $a_{\rm cr}\rightarrow 2.5$ and $\mucr\rightarrow 4.555$,
whereas $\psi_{\rm cr}\propto(4-3\gp)^{-1/4}$ grows
without bound.
One can readily show that equation (\ref{eq:mucr}) agrees with equation
(\ref{eq:musps}) for an SPS if the values of $a$ and $\psi$
for an SPS are inserted into this expression.  Note that
$\mucr\rightarrow 0$ as $\gp\rightarrow 0$: the critical mass
shrinks with $\gp$ since it becomes impossible to maintain
a substantial pressure gradient as $\gp$ approaches zero.

\subsection{Results for non--isentropic, 
locally adiabatic polytropes ($\g\neq\gp$)}
\label{sec:non}

\subsubsection{Stellar vs. pressure--confined polytropes}
\label{sec:stellar}

	The behavior of locally adiabatic polytropes is summarized
in Figure \ref{fig:laggb}. Since such
polytropes adhere to the Schwarzschild criterion for
stability against convection, those with $\g<\gp$ are
convectively unstable; we do not consider them further.
Locally adiabatic polytropes can be divided into two classes:

	{\it Stellar polytropes} ($\gamma> 4/3$, $\gp>6/5$)  
can serve as models for stable stars.
Polytropes with $\g>4/3$ are unconditionally
stable (Ledoux 1965), since
a locally adiabatic gas with $\g>4/3$ becomes hotter at every point under 
compression and therefore cannot collapse under its own gravity.
(Polytropes with $\gamma=4/3$ and zero pressure at their boundaries
are neutrally stable.)
Furthermore, polytropes with 
$\gp>6/5$ can have a zero-pressure boundary at finite
mass and radius, like stars (\cite{cha39}).
(Recall that polytropes with $\gp>6/5$ 
can have all their energy concentrated in the core---\S 
\ref{sec:singular}.)  
As the surface pressure approaches zero ( $\chi\rightarrow\infty$)
the dimensionless mass
$\mu\rightarrow 0$ for $6/5<\gp<4/3$, 4.555 for $\gp=4/3$,
and $\infty$ for $\gp>4/3$, as can be inferred from
Chandrasekhar (1939).  For fixed $\gamma$, there is
no limit to the dimensional mass of a stellar polytrope; stars
have limited masses because $\gamma$ depends on
the mass, approaching 4/3 for massive, radiation--dominated
stars or for degenerate stars that approach the Chandrasekhar
limit.

        {\it Pressure--confined polytropes} ($\gamma\leq 4/3\ and/or\  
0<\gp\leq 6/5$):  First consider
polytropes with $\gp<6/5$.  All such polytropes must be confined by
the pressure of an ambient medium (Chandrasekhar 1939).  In the
limit of a large center--to--surface density ratio ($\chi\gg 1$),
the pressure at the surface of such a polytrope approaches that of an SPS,
\beq
P_s=\frac{\gp(4-3\gp)c_\s^4}{2\pi G(2-\gp^2)^2r^2}.
\label{eq:ps}
\eeq
(Polytropes with $\gp=6/5$ do not satisfy this relation
because their mass and energy are concentrated at the center,
but they too must be pressure--confined if they have a finite
radius.)  Next, consider polytropes with $4/3\geq
\gp\geq 6/5$ and $\gamma<4/3$.
Such polytropes can have arbitrarily
large values of $\chi$, but
they become unstable
for $\chi>\chicr$; as a result, {\it stable} polytropes
in this region of parameter space must also be pressure--confined.
The mass and energy of these polytropes is concentrated near
the center when $\chi$ becomes large.
Finally, polytropes with $\gamma=4/3$ must be pressure--confined
if they are to be stable (and not just neutrally stable)
against small perturbations.
Polytropic models for thermal pressure in interstellar gas
and for magnetic
pressure are in the pressure--confined regime, and
as we shall see below, polytropic models for Alfv\'en waves
are equivalent to locally adiabatic polytropes in this regime.
Insofar as the pressure in interstellar clouds 
can be represented as the sum of these pressure components,
it follows that polytropes used
to model interstellar clouds must be pressure--confined.

\subsubsection{Limitations on the mass and radius
of pressure--confined polytropes}
\label{sec:limitations}

	The parameters $\mucr$ and  $\mucr'$, which describe 
the critical mass of a cloud, are plotted in Figures {\ref{fig:mucr}
and {\ref{fig:mucrp}.
Recall that for the isentropic case, $\mucr$ is a local
maximum in $\mu(\chi)$ for $\gp<4/3$ (\S \ref{sec:results1}); 
numerical solutions show that this is a global maximum as well, so
that $\mucr({\rm non-isentropic})<\mucr({\rm isentropic})$.  Furthermore,
as indicated in Figure \ref{fig:mucr}, 
$\mucr({\rm isentropic})$ is a monotonically increasing
function of $\gp$, reaching $\mucr=4.555$ at
$\gp=4/3$.  
The maximum value of $\mucr'$ generally occurs for  non--isentropic
clouds, and so cannot be read off Figure \ref{fig:mucrp}.  
We find that it too is a monotonically increasing function
of $\gp$, reaching max($\mucr')=1.686$ at $\gp=4/3$.
We conclude that all polytropes with
$\gp\leq 4/3$,
whether isentropic or not, satisfy $\mu\leq 4.555$
and $\mu'\leq 1.686$, so that
\beq
M\leq 4.555\left[\frac{c_\s^3}{(G^3\rho_\s)^{1/2}}\right],\ \ \ 
M\leq 1.686\left[\frac{\avg{c^2}^{3/2}}{(G^3\rho_\s)^{1/2}}\right].
\eeq
In terms of the surface pressure instead of surface density, we
find
\beq
M\leq 4.555\left[\frac{c_\s^4}{(G^3P_\s)^{1/2}}\right],\ \ \ 
M\leq 1.409\left[\frac{\avg{c^2}^2}{(G^3P_\s)^{1/2}}\right].
\eeq
If attention is restricted to
negative--index polytropes, the same line of 
reasoning that led to an upper limit of
4.555 on $\mu$ for $\gp<4/3$
leads to an upper limit $\mu\leq\mucr<\mucr(\gamma
=\gp=1)=1.1822$ for $\gp<1$.  Our results show that for such polytropes
$\mu'\leq {\rm max}(\mucr')<1.1822$ as well.
Note that since
$\mucr$ decreases below $\mucr({\rm isentropic})$ 
as $\gamma$ increases
above $\gp$, it is possible for
$\mu$ to exceed $\mucr$ for stable clouds (similarly,
$\mu'$ can exceed $\mucr'$), but this is by less than
a factor 1.5 for $\gp\leq 1$.  

	 Our results also set limits on the size of polytropes.
As indicated in Figure \ref{fig:ml}, the maximum value
of $\lambda$ occurs prior to the critical point.
This maximum value increases monotonically with $\gp$
for $\gp\leq 4/3$,
reaching 0.6603 at $\gp=4/3$.  Therefore,
all polytropes with $\gp\leq 4/3$ satisfy
\beq
r\leq 0.6603\left[\frac{c_\s}{(G\rho_\s)^{1/2}}\right]=
      0.6603\left[\frac{c_\s^2}{(GP_\s)^{1/2}}\right].
\eeq
For negative index polytropes, the upper limit on
the radius has the same form, but the coefficient is reduced
to 0.5142.

\subsubsection{Pressure and density drops in non--isentropic polytropes}
\label{sec:pressure}

	As discussed in the Introduction, an important question
to be addressed by the polytropic models is whether they
can account for the observed densities and pressures
in molecular clouds.  First consider the {\it mean}
densities and pressures.  These quantities have been
plotted for isentropic clouds in Figure \ref{fig:rhobar},
and they diverge as $\gp\rightarrow 4/3$.  For
$6/5\leq\gp<4/3$, the mean--to--surface density
and pressure ratios can become arbitrarily large
as $\gamma\rightarrow 4/3$, since such polytropes
have their mass and energy concentrated near the center,
as discussed above (\S \ref{sec:stellar}).
For $\gp<6/5$, the mean--to--surface density
and pressure ratios can no longer diverge.
For negative index polytropes ($\gp<1$),
it is convenient to express our results for the mean
densities and pressures in terms of the values for SPSs
(eqs. \ref{eq:rhobar}, \ref{eq:pbar}), 
\beqa
\frac{\bar\rho}{\rho_\s} & \leq & 1.26\;\frac{3(2-\gp)}{(4-3\gp)},\\
\frac{\bar P}{P_\s} & \leq & 1.26\;\frac{3(2-\gp)}{(6-5\gp)}.
\eeqa
We conclude that {\it negative--index polytropes cannot
have large mean--to--surface density or pressure ratios,
regardless of the value of the adiabatic index $\gamma$}.

	   Next, consider the 
{\it central}--to surface
density and pressure ratios, which 
can become quite large for non--isentropic clouds.
For fixed $\gp$, the maximum
value of $\chi$ for a stable equilibrium (i.e., $\chicr$)
increases as $\gamma$ increases above $\gp$, and it is
possible for $\chicr$ to
reach infinity if $\gamma$ exceeds a value
we label $\gamma_\infty$.
We can determine the value of
$\gamma_\infty$ analytically by noting that
the cloud becomes a singular polytropic
sphere with constant values of $\mu$ and $\lambda$ as
$\gamma\rightarrow\gamma_\infty$.
After a little algebra, we find that the requirement
that the critical point ($\dlm\rightarrow 0$)
occur in the limit $\chi\rightarrow\infty$
implies
\beq
\gamma_\infty=\frac{32\gp(2-\gp)}{(6-\gp)^2}~~~~~~~~
	(\gp\leq\frac 65).
\label{eq:ginfty}
\eeq
For $\gp=1$, this yields $\gamma_\infty=32/25=1.28$.
Yabushita (1968) previously pointed out that spatially
isothermal spheres with $\chi=\infty$ are unstable
for $\gamma<32/25$.  For $\gp=6/5$, $\gamma_\infty=4/3$;
for $4/3\geq\gp>6/5$, $\gamma_\infty$ remains at 4/3. 
Note that polytropes with $\gamma>\gamma_\infty$
and $\gp<6/5$ satisfy
equation (\ref{eq:ps}) for the surface pressure, 
so the infinite pressure drop is attained only at 
infinite radius.  Since actual clouds are finite in extent,
they are pressure--confined as discussed above.

     There is a subtlety in the behavior
of non--isentropic clouds with large density drops, 
however.  Recall that
as $\chi\rightarrow\chicr$, the pressure perturbation
$\delta\ln P_\s\rightarrow 0$ according to the condition
for the critical point (\S \ref{sec:general}).  
For $\gp=1.2$, for example,
the pressure perturbation 
within the sphere remains
comparable to the central value $\delta\ln P_{\rm c}$
throughout most
of the volume and approaches 0 only near the edge.
For negative--index polytropes, though, the
pressure perturbation can drop to very small
values well inside the sphere.
Consider the example of a sphere with
$\gp=0.5$ and $\gamma=0.79$ (slightly
smaller than $\gamma_\infty=0.793$).  We find that
$\delta \ln P$ drops to $0.1\delta \ln P_{\rm c}$ at
$\chi\sim 4$, 
far smaller than $\chicr\simeq 32$.  Thus, 
for any such polytrope with $\chi>4$, the
surface perturbations must be very small in order
to prevent the perturbations at the center from
becoming nonlinear.  
Although our linear analysis cannot determine what would
occur if the central pressure perturbation became
nonlinear, it is quite possible that the cloud would
become unstable.  If so, the
stability of the cloud in this case could be likened
to that of a pencil balanced on a flattened point---as
$\chi$ increases, the
size of the flattened region shrinks so that
it becomes susceptible to smaller and smaller
perturbations, and as a result, such a cloud would be unlikely
to survive in a medium with large pressure fluctuations.
As $\gamma$ increases, the value of $\chi$ at
which the pressure fluctuation becomes substantially
smaller than the central value grows.
We find that a sufficient condition
for density drops of at least $10^3$
to occur in clouds with 
$\delta \ln P_\s/\delta \ln P_{\rm c}>0.1$
is $\gamma >1.3\gp^{1/2}$.  For
$\gp<0.8$, clouds with $\gamma$ just above
$\gamma_\infty$ do not satisfy this condition
even though they are formally stable for arbitrarily large
density drops.  It is interesting to note that
insofar as logatropes can be represented by polytropes
with $\gp\rightarrow 0$, they do become stable for
very large density drops provided $\gamma$ is sufficiently
large; the central temperature is not constant unless
$\gamma=1$, however.

      Finally, we consider the response of locally adiabatic
polytropes to pressure perturbations.
For isothermal polytropes, $\Mcr$ is reduced
somewhat by an increase in the external pressure, 
$\Mcr\propto P_s^{-1/2}$ (eq. \ref{eq:Mcr}).  
For negative--index polytropes, however,
$\Mcr$ can decrease much more sharply with $P_\s$
due to the decrease in $c_\s$ (\cite{shu72}, \cite{toh87}).
Since the decrease in the temperature is bounded
(it is difficult to cool below 10 K in a typical
molecular cloud, for example), it is more convenient
to express the critical mass in terms of quantities
after the compression,
\beq
\Mcr=\mucr\left(\frac{c_{\s,f}^4}{G^{3/2} P_{\s,f}^{1/2}} \right) ,
\label{eq:mcr3}
\eeq
where $c_{\s,f}$ is the final value of $c_\s$, etc.
This result shows that the 
reduction in the critical mass due to cooling,
which reduces $c_\s$, can be large, but the reduction due to
the compression is limited by the weak $P_{\s,f}^{-1/2}$
dependence.  For example, 
in a radiative shock the final pressure 
is related to the initial value $P_{\s,i}$ by
$P_{\s,f}=(v_{\rm shock}/c_{\s,i})^2 P_{\s,i}$.
In spherical implosions even higher compressions,
and correspondingly greater reductions in $\Mcr$,
are possible (\cite{toh87}), although in practice
it may be difficult to maintain the high degree
of spherical symmetry required to achieve very 
large compressions.

\section{Globally Adiabatic Polytropes}

	When the flow of energy is significant on a
dynamical time scale, the assumption that the
perturbation is locally adiabatic breaks down.
The perturbation can be approximated as being globally adiabatic
provided that no heat is supplied to or removed from
the cloud
during the perturbation.  This is the 
approximation that we adopt for the Alfv\'en  waves
we are using as a model for the turbulence in
molecular clouds.  Precisely because heat can flow
during a perturbation, determining the condition
that the perturbation be adiabatic is non-trivial,
and it is this problem we now address.

\subsection{Entropy of a Globally Adiabatic Pressure Component}
\label{sec:entropy2}

	For a globally adiabatic pressure component
$g$, heat is assumed to flow
within the system
so as to maintain the polytropic condition $P_g=\kpg \rho^\gpg$,
but there is no exchange of heat with the surroundings,
nor with the other pressure components.
As a result of the heat flow, $\kpg$ may change with time, although
it is independent of position.
In the actual globally adiabatic system 
(as opposed to our model of the system), 
this heat flow
is reversible.  For example, if a non-uniform cloud with Alfv\'en
vwaves is adiabatically compressed and then decompressed back to its
original size, the Alfv\'en waves will be unchanged by the
process, since the wave action is conserved (Dewar 1970).
This leads to a problem: 
in our model for the Alfv\'en waves, heat flows
across a temperature gradient---an irreversible process.
As a result, the entropy as conventionally defined may not be conserved
for globally adiabatic processes.  

	To define an entropy for a globally adiabatic pressure
component, we again introduce a 
spherically symmetric reference state that is
related to the cloud by a reversible adiabatic 
transition (c.f. Mouschovias 1976a).  In this reference state, the
globally adiabatic pressure component has a uniform pressure
$P_{g,{\rm ref}}$; since globally adiabatic components
are assumed to always be polytropic, it follows that the
density $\rho_{g,{\rm ref}}$ is uniform as well. 
The entropy parameter for this component in the reference state is 
\beq
\kgr\equiv \frac{\pgref}{\rhoref^\gamg}.
\label{eq:kgr}
\eeq
If the cloud then returns to its initial state by a reversible
adiabatic process,  the entropy will be unchanged.  
Thus, the entropy of a globally adiabatic pressure component 
is characterized by
$\kgr$.  In our case,
we are interested in self-gravitating clouds, so the transition
to the reference state can be visualized as being due to slowly
decreasing the value of the gravitational constant $G$ to zero.

	In Appendix \ref{app:global}, we evaluate $\kgr$ by extending
an argument due to McKee \& Zweibel (1995).  There we
demonstrate that $\kgr=\kavg$, where $\kavg$ is
the mean entropy parameter for a globally adiabatic pressure
component, defined as
\beq
\kavg^{\left(\frac{\gpg-1}{\gpg-\gamg}\right)}\equiv\frac{1}{M}\int
	K_g^{\left(\frac{\gpg-1}{\gpg-\gamg}\right)} dM.
\label{eq:kavg}
\eeq
Observe that since 
\beq
K_g=C\exp[(\gamg-1) s_g/k],
\label{eq:kofs}
\eeq
where $C$ is a constant and $s_g$ is the specific entropy 
of a gas element (eq. \ref{eq:entropy}),
it follows that the entropy parameter for a globally adiabatic
pressure component is given by a well-defined integral
of the specific entropy.  However, in contrast to the case
for a locally adiabatic pressure component (\S 5.1),
the specific entropy $s_g$ may {\it not} be constant during
a globally adiabatic change, due to internal
heat flows; it is only the integral in equation
(\ref{eq:kavg}) that remains constant.

	For a spatially isothermal component, for which
there is no heat flow across a temperature gradient,
the entropy parameter is related to the mean
specific entropy $\avg{s_g}=M^{-1}\int s_g dM$ in a simple way.
Let $\epsilon\equiv \gpg-1$ be small. Inserting
$K_g(s_g)$ from equation (\ref{eq:kofs}) into
equation (\ref{eq:kavg}), we have
\beq
\kavg^{\frac{\epsilon}{\gpg-\gamg}}=\frac{C}{M}\int \exp\left[\epsilon
	\left(\frac{\gamg-1}{\gpg-\gamg}\right)\frac{s_g}{k}\right] dM.
\eeq
Expanding both sides of this equation for small
$\epsilon$, we find
\beq
\ln\kavg=\frac{\gamg-1}{k}\,\avg{s_g}+{\rm const}.
\eeq
Since $\kavg$  is constant for a spatially
isothermal, globally adiabatic system, it follows that
$\avg{s_g}$ is too (c.f. LBW).
However, if the system
is neither spatially isothermal 
nor isentropic (i.e., $\gpg\neq 1$ and $\gpg\neq\gamg$),
then only $\kavg$ is constant.

	The polytropic constant $\kpg$ evolves during
a globally adiabatic change as
\beq
\kpg^{\gamg-1}=\kavg^{\gpg-1}\cavgt^{\gamg-\gpg}
\label{eq:kpg1}
\eeq
from equation (\ref{eq:kgr2}).
This can also be expressed in terms of the density
by noting that $\cg^2=\kpg\rho^{\gpg-1}$, so that
\beq
\kpg=\kavg\avg{\rho^{\gpg-1}}^{{\gamg-\gpg}\over{\gpg-1}}.
\label{eq:kpg2}
\eeq
This equation allows one to follow the evolution of the pressure 
of a globally adiabatic component, $P_g=\kpg \rho^\gpg$, 
in terms the evolution
of the density structure of the cloud.

\subsection{The Critical Mass}
\label{sec:critical2}

	To determine the critical point for a cloud supported
by a globally adiabatic pressure component,
we express the mass as
\beq
M=\frac{\mu P_\s^{3/2}}{G^{3/2}\rho_\s^{2}}
	=\frac{\mu P_\s^{(3\gpg-4)/2\gpg}\kpg^{2/\gpg}}{G^{3/2}}.
\eeq
Using this equation to evaluate the effect
of a pressure variation, we find that 
the critical point ($\delta P_\s=0$) occurs at
\beq
\delta\ln\mu+\left(\frac{2}{\gpg}\right)\dlkg=0,
\eeq
which is just what we found in 
equation (\ref{eq:critpt}).
Determination of the location
of the critical point requires evaluation of
$\dlkg$ at constant entropy, which is carried out in Appendix 
\ref{app:global}.

\subsection{Results for Globally Adiabatic Clouds}
\label{sec:results2}

	There are two key differences between the stability
of globally adiabatic clouds and that of locally adiabatic
clouds: First, there is no convective instability, since
it is assumed that the heat flow is rapid compared to fluid
motions; as a result, it is possible to have stable
systems with $\gamg<\gpg$, although we have not identified
one in practice.  Second, as discussed in \S 4.1.2,
clouds with $\gamg>4/3$, which are stable if they
are locally adiabatic, can be unstable to core collapse.
In core collapse, instability is
triggered when a perturbation causes a sufficiently
large heat flow from the center of the cloud to the envelope.
This is the case that is relevant to clouds supported
by Alfv\'en waves, which have $\gamw=3/2$
(see \S \ref{sec:clouds}).  Globally adiabatic polytropes with
$\gp>6/5$ and $\gamg>4/3$ are not subject to core collapse, however:
As shown in \S \ref{sec:non}, polytropes with $\gp>6/5$
can have arbitrarily large density contrasts, and
our numerical results show that they
are stable for $\gamg>4/3$.

	Results for the center--to--surface pressure
and density ratios, $P_{\rm c}/P_\s$ and $\rho_{\rm c}/\rho_\s$,
are plotted in Figures \ref{fig:gapr} and \ref{fig:gaden},
respectively.
Note that for the cases of greatest interest,
in which $\gamg>\gpg$, these ratios are determined
primarily by the value of $\gpg$.  
Comparing these results with those for
locally adiabatic polytropes discussed in \S \ref{sec:locally},
we find that globally adiabatic
polytropes have smaller critical density and
pressure contrasts
(provided $\gamma>\gp$, the only case of relevance
for locally adiabatic polytropes).

    It is important to note that the structure of 
a cloud supported by a globally adiabatic pressure component
is determined by the value of $\gp$ and is therefore
identical to that of a cloud supported by a locally adiabatic
pressure component with the same value of $\gp$.  
For a locally adiabatic cloud, the density drop 
at the critical point $(\rhoc/\rhos)_{\rm cr}$
rises smoothly from the isentropic value (Fig. \ref{fig:rho})
to $\infty$ as $\gamma_\ell$ rises from
$\gpl$ to $\gamma_\infty$.  Figure \ref{fig:gaden} shows that
$(\rhoc/\rhos)_{\rm cr}$ increases above
the isentropic value as $\gamma$ increases above
$\gp$ for a globally adiabatic component as well.
Hence, for any globally adiabatic cloud with $\gamg>\gp$, it 
is always possible to find a value of $\gl$, which we label
$\gl$(equiv), such that the critical point of the locally
adiabatic cloud is identical to that of the globally adiabatic cloud.
The value of $\gl$(equiv) for the equivalent locally adiabatic
cloud is in the range $\gamma_\infty>\gl$(equiv)$>\gp$.  
We have evaluated $\gl$(equiv)
for several cases:  
For $\gamma=3/2$ and $\gp=0.5$ (the Alfv\'en wave case--see below), 
$\gl$(equiv)$=0.60564$;
for $\gamma=3/2$ and $\gp=0.1$, $\gl$(equiv)$=0.11985$;
and for $\gamma=5/3$ and $\gp=1.0$ (the LBW case), 
$\gl$(equiv)$=1.22042$.  It must be kept in mind that although
it is possible to locate the critical point 
in a globally adiabatic polytrope with a
locally adiabatic equivalent, the nature of
the instability in the two cases is quite different: for
$\gamg>4/3$, the globally adiabatic polytrope is subject to
core collapse, which is impossible for a locally adiabatic polytrope.

     How does this result carry over to the case of a cloud
with multiple pressure components?  Just as in the single
component case, the structure of the cloud is determined
by the values of $\gp$ for each of the components,
with the values of $\gl$ and $\gamg$ determining which of the structures
corresponds to the critical point.  The critical point
for a cloud supported by 
one or more locally adiabatic pressure components plus
a globally adiabatic pressure component is therefore
the same as that of a cloud supported by the same locally adiabatic
components plus a locally adiabatic component with a suitable
$\gl$(equiv).  Numerical calculations for two component clouds
show that the
value of $\gl$(equiv) is not constant, but instead
depends on the fraction of the pressure in the globally adiabatic
component.  This is reasonable, since although the structure
equations do not depend on the values of $\gl$ and $\gamg$, the
equations for the perturbed structure do (\S \ref{sec:equations}).
In contrast to the single component case, the value of
$\gl$(equiv) can exceed $\gamma_\infty$, although for Alfv\'en
waves it is generally $\simlt 1$.

\subsection{Clouds Supported by Alfv\'en Waves}
\label{sec:clouds}

	We now focus on Alfv\'en waves, which we
are using to model the turbulence in molecular clouds.
As shown by Dewar (1970), Alfv\'en waves exert
an isotropic pressure and can thus provide
support parallel to the background magnetic field.
For small  amplitude waves, McKee \& Zweibel (1995)
showed that $\gamw=3/2$ and $\gpw=1/2$; this latter
result was originally derived by Wal\'en (1944).
While Alfv\'en waves are undoubtedly an important component
of the MHD waves that contribute to the turbulent pressure
in molecular clouds, there are several limitations to
our model that should be borne in mind.  First,
we assume that the waves are adiabatic, whereas
simulations suggest that the waves damp rapidly
(\cite{vaz99}).  Second, following Dewar,
we assume that the waves are in the WKB limit
(short wavelength), whereas the observations suggest
that the largest amplitudes are in waves with wavelengths
comparable to the size of the cloud; as a result, we
expect our models to be more accurate in reproducing the 
mean value of the wave pressure than the value of the
wave pressure at the edge of the cloud, for example.  Finally,
whereas simulations of 1D
MHD turbulence by Gammie \& Ostriker (1996)
confirmed that $\gpw=0.5$ for small
amplitudes, they indicate that $\gpw$ becomes significantly smaller
for large amplitudes.  It is not clear that this
result can be applied to GMCs, however: much of
the reduction in $\gpw$ they found was due to the
fact that the gas became highly clumped at large amplitude,
and the wave pressure was not much larger inside the clumps
than outside.  However, the question for GMCs is whether
Alfv\'en waves can maintain a pressure gradient in a clumpy
medium, and this would require a much larger scale simulation.

	It is possible to infer $\gpw$ from observation,
but the existing results are not definitive.
If observations show that the density
scales as $\rho\propto r^{-p}$ and the 1D nonthermal velocity
dispersion scales as
$\snt\propto r^q$, then $P_{\rm w}\propto \rho\snt^2$ implies
\beq
\gpw=1-\frac{2q}{p}.
\eeq
Myers' (1985) reanalysis of Larson's (1981) heterogeneous data
gave $p=1.2$ and $q=0.3$, corresponding to $\gpw=0.5$.
When Caselli
\& Myers (1995) focused on low mass cores, they found
$p=1.1$ and $q=0.53$, corresponding to $\gpw\simeq 0$;
such cores are believed to be largely supported by
static magnetic fields and thermal pressure, however.
(It should also be noted that their value of $p$ applies
well outside the inner, isothermal core.)
For massive cores, where the non-thermal motions 
are more significant, they found $p=1.6$ and $q=0.21$,
corresponding to $\gpw\simeq 0.75$.  With the exception
of low mass cores, then, it appears that $\gpw\sim 0.5$
is not inconsistent with the data.

	Adopting $\gpw=0.5$ and $\gamw=1.5$, we find
that the maximum possible pressure
ratio for a cloud supported by Alfv\'en waves is only 4.15;
the corresponding density ratio of 17.26 is somewhat greater than
that for a critical isothermal cloud.
The polytropic constant varies as
$\kpg=\cavgt^2/\kavg=\kavg\avg{\rho^{-1/2}}^{-2}$
(eqs. \ref{eq:kpg1} and \ref{eq:kpg2}).
The critical mass for a cloud supported by Alfv\'en waves
can be expressed in terms of the non-thermal velocity dispersion
$\snt$ since the Alfv\'en wave pressure,
including
the pressure of the fluctuating fields, is
\beq
P_{\rm w}\equiv \rho c_{\rm w}^2=\frac 32 \rho\snt^2
\label{eq:pw}
\eeq
(McKee \& Zweibel 1995).
For $\gamw=3/2$ and $\gpw=1/2$,
we find $\mucr=0.21334$, 
$\psi_{\rm cr}=(\cavgt/c^2)_{\rm cr}=(\avg{\snt^2}/\snt^2)_{\rm cr}=
0.71342$, and $\mucr'=0.35405$.  The critical mass in this case is
then
\beq
\Mw=0.3919\;\frac{\snt^2}{G^{3/2}\rho_\s^{1/2}}=
	0.6504\;\frac{\avg{\snt^2}^{3/2}}{G^{3/2}\rho_\s^{1/2}}.
\label{eq:Mw}
\eeq
This result for the critical mass 
is less than the estimate of McKee \& Zweibel (1992)
based on the virial theorem; they found a coefficient
of 2.17 in the second expression.  This discrepancy is due to the fact that
the maximum possible value of $\mu$ in equilibrium 
(independent of stability) is governed by
$\gp$, which does not enter the virial theorem argument.
In inferring the maximum mass allowed by the virial theorem,
one implicitly assumes that the three terms 
in equation (\ref{eq:virial}) can be of similar magnitude;
however, for values of $\gp$ significantly less than 1, 
the pressure becomes approximately constant in the cloud
and this is not allowed.

	Although the results of this paper are primarily
for single component polytropes, we shall present two
results for multi--pressure polytropes.  We find that
the critical mass for a cloud consisting
of a cloud supported by isothermal gas pressure and
Alfv\'en waves is approximately
\beqa
\Mcr&=&(\Mbe^{2/3}+\Mw^{2/3})^{3/2},\\
    &=&1.182\left(\frac{\seff^3}{G^{3/2}\rho_\s^{1/2}}\right),
\label{eq:mcriso}
\eeqa
where the effective velocity dispersion
$\seff$ is given by 
\beq
\seff^2=\cth^2+(\mu'_{\rm w,cr}/\mu'_{\rm BE,cr})
	^{2/3}\avg{c_{\rm w}^2}.
\label{eq:seff1}
\eeq
For Alfv\'en waves with $\gp=1/2$, the effective velocity dispersion 
becomes
\beq
\seff^2\equiv \cth^2+0.67\avg{\snt^2};
\label{eq:seff2}
\eeq
equation (\ref{eq:mcriso}) is accurate to within 2\% in this case.
Thus Alfv\'en waves are less effective at supporting
clouds despite the fact that their pressure is 
1.5 times larger than $\rho\snt^2$ due to the fluctuating magnetic
fields.

	Second, in Figure \ref{fig:mpp}
we show the density distribution in a multi--pressure
polytrope that reproduces some of the features observed in molecular
clouds.  We have chosen the polytropic
index of the magnetic pressure to be $\gpb=1.0$ in order to
illustrate the dramatic effect of a non--isentropic
pressure component (recall that $\gamma_B=4/3$).  
The magnetic pressure is comparable with the Alfv\'en wave
pressure, and the thermal pressure is 10\% of the 
magnetic pressure.
The case shown is a subcritical cloud, with
a center--to--surface density contrast of 545
compared to the critical value of 1737; this density drop
is far larger than that of the Bonnor--Ebert sphere shown
for comparison, which has a density drop of 14.  
On the other hand, the mean--to--surface density
contrast of 2.59 for the multi--pressure polytrope 
is only slightly greater than the value of 2.46 for
the Bonnor--Ebert sphere.  The inclusion
of the wave pressure is destablizing: in the absence of
wave pressure, a cloud supported by a
magnetic pressure with $\gpb=1.0$ and a thermal pressure
$\pth=0.1P_B$ is stable
for an arbitrarily large density drop.

\section{Summary}
	
	The structure of molecular clouds has a major effect
on the star formation that occurs within them.
Polytropic models, in which the pressure varies as a power
of the density ($P\propto \rho^\gp$), are useful in the study of
the structure of such clouds just as they are for the study
of stellar structure
(Chandrasekhar 1939).  Like stars, molecular clouds 
and some of the clumps within them are
self-gravitating gas clouds, but there are two important 
differences:
molecular clouds have a soft equation of
state (for example, the gas is often approximately isothermal),
so that they must be confined by an 
external pressure;
and they are highly turbulent, which leads to a complex internal
structure and a non-spherical shape, both of which are time
dependent.  In order to determine the 
characteristics of stable molecular clouds and the 
conditions under
which molecular clouds become unstable to gravitational collapse,
we have assumed that the pressure in these clouds
can be represented by a sum of polytropic pressures---a 
multi--pressure polytrope.
The individual terms represent thermal gas pressure,
magnetic pressure and turbulent pressure.
We model the turbulent pressure as being due to
a superposition of Alfv\'en waves, which
have a polytropic index $\gp= 1/2$.
This is an example of a ``negative--index polytrope"
(so called because the index $n$ in $\gp=1+1/n$ is negative),
which has long been known to account for the
outward increase in line widths, both thermal
(Shu et al 1972) and non--thermal (Maloney 1988).
Our models assume hydrostatic equilibrium and 
spherical symmetry, both of which become better approximations
if a time average is taken.  Rotation is neglected, which appears
to be a reasonably good approximation (Goodman et al 1993).

	Several features of our analysis bear mention.
First, since interstellar clouds are confined by the pressure
of the ambient medium,
we describe the cloud in terms of variables 
($\mu\equiv M[G^{3/2}\rho^{1/2}/c^3]$ and $\lambda\equiv
r[G^{1/2}\rho^{1/2}/c]$, where $c\equiv [P/\rho]^{1/2}$)
that depend only on local properties of the cloud,
not on the properties at the center of
the cloud.  The variable $\mu$ is proportional to the ratio
of the mass to the local value of the Jeans mass,
whereas $\lambda$ is proportional to
the ratio of the radius to the local value of the Jeans radius.
The resulting description is simpler than the
standard one (see Appendix \ref{app:a} and Stahler 1983).  
Next, we assume that the perturbations are
adiabatic.  As is often done
in the study of stellar structure, we allow the adiabatic
index $\gi$, which measures how the pressure
of the $i$th component
responds to a perturbation in density, to differ from the polytropic index
$\gpi$.  The structure of the cloud depends on the values of
$\gpi$, whereas the stability depends on the values of $\gi$ as well.
Third, for pressure components that are not isentropic (i.e., those for
which $\gi\neq\gpi$), we consider two cases: For a locally adiabatic
component, such as the thermal pressure or the magnetic pressure,
the entropy of each mass element remains constant
during a perturbation, but as a result the perturbed cloud
is no longer a polytrope.  On the other hand,
for a globally adiabatic component, the entropy flows 
so as to maintain the polytropic relation between the pressure
and density.  This case is appropriate for Alfv\'en waves
under the assumptions that the perturbation is applied quasistatically
and that there are no sources or sinks for the waves.
We describe both locally and globally adiabatic pressure
components in terms of the entropy
parameter $K_i=P/\rho^\gi$.  For a thermal gas, $K_i$ is just
$\cth^2$; for the magnetic field, the appropriate average
of $K_i$ is related to the magnetic flux; and for Alfv\'en waves,
it is related to the wave action.  In Appendix \ref{app:global}, we show
how to define the mean entropy parameter for a globally
adiabatic pressure component so that it remains
constant during an adiabatic change.

	 The essential feature
of the three pressure components that support molecular clouds
is that their equations of state are 
{\it soft}, with a limited ability to resist gravity:
The thermal gas is approximately isothermal ($\gp\simeq\gamma\simeq 1$)
in regions of high extinction.
The magnetic field must have $\gamma=4/3$ in order to 
ensure flux freezing in spherical symmetry (\S \ref{sec:entropy1}); 
stability against the interchange instability requires
$\gp\leq\gamma= 4/3$. 
The turbulent pressure is modeled with
Alfv\'en waves, which contribute a significant
fraction of the pressure in MHD turbulence, and these waves 
can be rigorously
shown to obey $\gamma=3/2$ and $\gp=1/2$ in the limit
of small amplitude, small wavelength,
and negligible damping (McKee \& Zweibel 1995). 
In contrast to the gas and the field, Alfv\'en waves
are globally adiabatic, and clouds supported by
Alfv\'en waves are subject to core collapse if
the pressure drop inside the clouds is too large. 
As discussed in \S \ref{sec:results2}, Alfv\'en waves
are equivalent to a locally adiabatic pressure component
with $\gl({\rm equiv})\simlt 1$ insofar as determining the conditions
for gravitational collapse.

	The results in this paper are primarily for polytropes
with only one pressure component; results for multi--pressure
polytropes are deferred to another paper.  We extend previous
work on polytropes with a locally adiabatic pressure to the
non-isentropic case ($\gamma\neq\gp$).  
The parameter space for the stability of 
locally adiabatic polytropes 
is portrayed in Figure \ref{fig:laggb}.
Clouds with $\gamma<\gp$ are unstable to convection
and are not considered here.  
Locally adiabatic polytropes are classified either
as stellar polytropes ($\gamma>4/3$ and $\gp>6/5$)
or as pressure--confined polytropes.  Stellar polytropes
can serve as models for stars since they
are not subject to gravitational collapse and
they can have a zero pressure boundary at a finite radius.
On the other hand, pressure--confined polytropes
must have a finite pressure at their surfaces,
either because $\gp\leq 6/5$, or in order to
be stable against gravitational collapse with
$\gamma\leq 4/3$.  A polytrope supported by any
of the pressure
components in a molecular cloud must be pressure--confined:
In contrast to the case of
a star, the stability of a molecular cloud is determined by 
conditions at its {\it surface}.

	We introduced the concept of a globally adiabatic
pressure component as an idealized model for treating the
turbulent motions in molecular clouds.  
Molecular clouds are magnetized, and as a result the
turbulent motions can be considered to be a superposition
of MHD waves.
The pressure associated with wave motions cannot be locally
adiabatic since the waves can move from one part
of a cloud to another under the influence of a perturbation.
However, in the absence of sources and sinks, the total 
entropy of the waves remains constant---the system
is globally adiabatic.
A gravitationally bound cluster of stars is another
example of a globally adiabatic system.
We have determined the appropriate entropy parameter for
a non-uniform, globally adiabatic pressure component;
for the spatially isothermal case ($\gp=1$), it is just
the exponential of the usual entropy evaluated by LBW.
We then generalized the standard treatment of the stability
of polytropes to include a globally adiabatic component with an
arbitrary value of $\gp$.
In contrast to locally adiabatic systems, globally adiabatic
systems can be unstable to gravitational collapse even
when $\gamma>4/3$; provided $\gp<6/5$, they undergo
core collapse beyond the critical point (the ``gravo-thermal
catastrophe" of LBW).  This is in contrast to the 
phenomenological model used by Maloney (1988) in which 
gravitational instability is impossible.
Our results for the critical
pressure and density ratios are shown in Figures \ref{fig:gapr}
and \ref{fig:gaden}; these are smaller
than for the corresponding locally adiabatic cases.  
Alfv\'en waves, which we have used to model for
the turbulent pressure, are particularly simple
because their pressure is isotropic; as shown in equations
(\ref{eq:mcriso}) and (\ref{eq:seff2}) and as discussed at
the end of \ref{sec:clouds}, the non-thermal motions
associated with such waves are somewhat less effective
in supporting a cloud than are the thermal motions of an
isothermal gas.  

	   The critical
mass---i.e., the maximum mass that is stable against gravitational
collapse for a given entropy distribution and
ambient pressure---is
evaluated in terms of the parameters $\mucr$ and $\mucr'$,
\beq
\Mcr=\frac{\mucr c_\s^3}{G^{3/2}\rho_\s^{1/2}}\;=\;
	\frac{\mucr'\avg{c^2}^{3/2}}{G^{3/2}\rho_\s^{1/2}}.
\eeq
The values of $\mucr$ and $\mucr'$ 
for isentropic polytropes
are plotted in Figures \ref{fig:mucr} and \ref{fig:mucrp};
the values of $\mu$ and $\mu'$ for singular polytropic
spheres are also shown.
The values of $\mucr$ and $\mucr'$ depend primarily on the value
of $\gp$; both are of order unity for $\gp\sim 1$ and
approach zero as $\gp\rightarrow 0$.  
The maximum value of $\mu$ for any polytrope with $\gp\leq 4/3$
is 4.555, the value of $\mucr$ at $\gp=4/3$.  Negative--index
polytropes have $\mu<1.1822$, the value of $\mucr$ for
an isothermal sphere.  The ratio of the {\it mean} density
or pressure to the surface value is less than 4 for any
negative--index polytrope.
On the other hand, the
density and pressure contrasts between the center and the edge
of a critically stable polytrope rise from the values shown
in Figure \ref{fig:rho}  as $\gamma$ becomes greater than $\gp$.
In the region denoted ``SPS" in Figure \ref{fig:laggb},
locally adiabatic, singular polytropic spheres are formally stable against
collapse; the region in which spheres with large density
contrasts are stable against finite perturbations may
be more limited, as discussed in \S \ref{sec:non}.
Non--isentropic polytropes may thus provide an explanation
for the very large density contrasts observed between 
cores within molecular clouds and the edges of the clouds
(see Fig. \ref{fig:mpp} and Curry \& McKee 1999).

\acknowledgments
We thank Frank Bertoldi, Lars Bildsten, Charles Curry,
Donald Lynden-Bell, Chris Matzner and Dean McLaughlin for helpful comments.
JHH acknowledges support from a grant from the LLNL branch of
the IGPP.
The research of CFM is supported by NSF grant AST95-30480
and a grant from the Guggenheim Foundation; his research on
star formation is supported in part by a NASA grant to the Center of
Star Formation Studies.

\bigskip
\appendix

\centerline{\bf APPENDIX}
\section{Standard Nondimensional Variables and Equations}
\label{app:a}

	A set of `standard' dimensionless variables and associated equations 
is generally used to describe
the structure of polytropic gas spheres ({\it e.g.,} \cite{cha39}).
Polytropes satisfy
\beq
P(r)=K_{\rm p}\rho(r)^\gp,
\label{eq:polytrope}
\eeq
where $K_{\rm p}$ is independent of position.
The standard equations for isothermal spheres
(with $\gp=1$) are different from those for
other polytropic spheres. An advantage of the variables we
use (or of the homology variables $u$ and $[n+1]v$) is
that a single set of equations applies to both cases.  
Furthermore, our formulation applies to multiple pressure components
as well, whereas the standard formulation is restricted to single
pressure components.
Here we compare the standard isothermal and polytropic variables 
and equations to those introduced in this work.
Some of these relations have been given previously by
Stahler (1983), who focused on the case $n>0$; his
$M_n$ and $R_n$ are proportional to our $\mu$ and $\lambda$,
respectively.

	In the standard equations, the polytropic constant $n$ describes the
structure.
The relationship between $n$ and the polytropic index $\gp$ is
$\gp=(n+1)/ n.$
Isothermal spheres have $\vert n\vert\to\infty$;
the case $n=-1$ corresponds to an isobaric cloud.

\subsection{Dimensionless Variables and Structure Equations}
\label{sec:Dimensionless}

	The standard dependent variables are:
\beq
\Psi\equiv\ln{\rho_{\rm c}\over\rho}
\mbox{~~~(isothermal~spheres)},
\label{eq:dependent isothermal}
\eeq
\beq
\theta\equiv\left(\rho\over\rho_{\rm c}\right)^{1/n}
\mbox{~~~(other~polytropes)},
\label{eq:dependent polytropic}
\eeq
where $\rho_{\rm c}$ is the central density.
In this work we introduce a \nondim\ mass $\mu$ in equation (\ref{eq:mu}) and a
\nondim\ radius $\lambda$ in equation (\ref{eq:lambda}) as dependent variables.

	The standard independent variable is a dimensionless radius:
\beq
\xi\equiv R\left[{1\over4\pi}{c^2\over G\rho_{\rm c}}\right]^{-1/2}
\mbox{~~~(isothermal~spheres)},
\label{eq:independent isothermal}
\eeq
\beq
\xi\equiv R\left[{\vert n+1\vert\over4\pi}
{c_{\rm c}^2\over G\rho_{\rm c}}\right]^{-1/2}
\mbox{~~~(other~polytropes)};
\label{eq:independent polytropic}
\eeq
the independent variable in our analysis is $\indep=\ln(\rho_{\rm c}/\rho)$.

	The differential structure equations are derived using the equation of
hydrostatic equilibrium (\ref{eq:HSE}).
In the standard nondimensional formulation, the equilibrium condition is
reflected in the Lane-Emden equation,
\beq
{1\over\xi^2}
{d\over d\xi}
\left(\xi^2{d\Psi\over d\xi}\right)=e^{-\Psi}
\mbox{~~~(isothermal~spheres)}
\label{eq:lane-emden isothermal}
\eeq
\beq
{1\over\xi^2}
{d\over d\xi}
\left(\xi^2{d\theta\over d\xi}\right)=
-{\vert n+1\vert\over n+1}\theta^n\mbox{~~~(other~polytropes)}.
\label{eq:lane-emden polytropic}
\eeq
The structure equations used in this work are equations (\ref{eq:mu_diff}),
(\ref{eq:lambda_diff}), and, for multi--pressure polytropes,
equation (\ref{eq:gamma_p_I}).

	The boundary conditions for integration are
$\Psi=0,~{d\Psi/ d\xi}=0$ at $\xi=0$ for
isothermal spheres, and
$\theta=1,~{d\theta/ d\xi}=0$ at $\xi=0$
for other polytropes.
In our formulation,
$\mu=0$ and $\lambda=0$ at $\indep=0$.
Note that both $\xi$ and $\indep$ vanish at the origin.

\subsection{Dimensional Quantities} \label{sec:Dimensional Quantities}

	Equation (\ref{eq:dependent polytropic}) 
states $\rho\propto\theta^n$;
together with the ideal gas law, this proportionality implies that
the temperature $T\propto\theta$, so that $c^2=c_{\rm c}^2\theta$.
Chandrasekhar (1939) and Viala and Horedt (1974) identify the following
additional proportionalities for polytropic structures:
$\xi\propto$~distance, $\theta'\propto$~temperature gradient,
$\theta^{n+1}\propto$~pressure,
and $(-\vert n+1\vert/[n+1])\xi^2\theta'\propto M(\xi)$.

	In the standard problem, the dimensional radius can be
determined using equations
(\ref{eq:independent isothermal}) and (\ref{eq:independent polytropic})
together with the equation of hydrostatic
equilibrium; alternate forms are:
\beq
R={GM\over c^2}\left[\xi{d\Psi\over d\xi}\right]^{-1}~~~
({\rm isothermal~spheres}),
\label{eq:r isothermal 2}
\eeq
\beq
R=-{1\over n+1}{GM\over c_{\rm c}^2}
\left[\xi{d\theta\over d\xi}\right]^{-1}~({\rm other~polytropes}),
\label{eq:r other 2}
\eeq
\beq
R={GM\over c^2}{\lambda\over\mu}~~~({\rm this~work}).
\label{eq:r general}
\eeq
We may write the pressure as:
\beq
P={1\over4\pi}{c^8\over M^2G^3}\xi^4
\left(d\Psi\over d\xi\right)^2e^{-\Psi}~({\rm isothermal~spheres}),
\label{eq:P isothermal 2}
\eeq
\beq
P={\vert n+1\vert^3\over4\pi}{c_{\rm c}^8\over M^2G^3}
\xi^4\theta^{n+1}\left(d\theta\over d\xi\right)^2~({\rm other~polytropes}),
\label{eq:P other 2}
\eeq
\beq
P={c^8\over M^2G^3}\mu^2~~~({\rm this~work}).
\label{eq:P general}
\eeq
We have chosen to express $R$ and $P$ in terms of $M(R)$ and
$c(R)$ for our variables, since in our approach the dependent
variables depend only on properties at the surface.  Alternatively,
one can express the results in our variables in terms of
the central sound speed; for a single component polytrope,
the required relation is $c^2=c_{\rm c}^2 \exp [-({\gp}-1)\indep]$.

	Finally, the mass can be expressed as:
\beq
M={1\over(4\pi)^{1/2}}{c^3\over G^{3/2}\rho_{\rm c}^{1/2}}
\xi^2{d\Psi\over d\xi}~~~({\rm isothermal~spheres}),
\label{eq:M isothermal}
\eeq
\beq
M=-{\vert n+1\vert^{5/2}\over(4\pi)^{1/2}(n+1)}
{c_{\rm c}^3\over G^{3/2}\rho_{\rm c}^{1/2}}\xi^2
{d\theta\over d\xi}~({\rm other~polytropes}),
\label{eq:M other}
\eeq
\beq
M={c^3\over G^{3/2}\rho^{1/2}}\,\mu ~~~({\rm this~work}).
\label{eq:M general}
\eeq
If desired, the last result can be expressed in terms of the central 
density by using the relation $\rho=\rho_{\rm c}\exp(-\indep)$.

\subsection{Conversion Among Dependent Variables}
\label{sec:Conversion Among Dependent Variables}
 
\noindent
{\it Isothermal Case:}

	Since $\Psi=\indep$,
we find
\beq
\mu={1\over(4\pi)^{1/2}}\xi^2{d\Psi\over d\xi}e^{-\Psi/2},
\label{eq:mu to standard isothermal}
\eeq
\beq
\lambda={1\over(4\pi)^{1/2}}\xi e^{-\Psi/2},
\label{eq:lambda to standard isothermal}
\eeq
and
\beq
\xi=(4\pi)^{1/2}\lambda e^{\indep/2},
\label{eq:xi to general}
\eeq
\beq
{d\Psi\over d\xi}={1\over(4\pi)^{1/2}}{\mu\over\lambda^2}e^{-\indep/2}.
\label{eq:d_Psi/d_xi to general}
\eeq

\noindent
{\it Polytropic Case:}
 
	Since $\theta=\exp(-\indep/n)$, 
we find
\beq
\mu=-{(n+1)\vert n+1\vert^{1/2}\over(4\pi)^{1/2}}
	\xi^2 \theta^{(n-3)/2}{d\theta\over d\xi},
\label{eq:mu to standard polytropic}
\eeq
\beq
\lambda={\vert n+1\vert^{1/2}\over(4\pi)^{1/2}}\xi\theta^{(n-1)/2},
\label{eq:lambda to standard polytropic}
\eeq
and
\beq
\xi={(4\pi)^{1/2}\over\vert n+1\vert^{1/2}}
\lambda e^{(2-\gp)\indep/2},
\label{eq:xi polytropic to general}
\eeq
\beq
{d\theta\over d\xi}=-{\vert n+1\vert^{1/2}\over(4\pi)^{1/2} (n+1)}
{\mu\over\lambda^2}e^{-\gp\indep/2}.
\label{eq:d_Psi/d_xi polytropic to general}
\eeq

\section{DERIVATION OF EQUATIONS FOR THE PERTURBED STRUCTURE}
\label{app:derivation}

\subsection{Perturbations in $\mu$ and $\lambda$}

	To assess the stability of multi--pressure polytropes,
we must determine the structure they assume after a perturbation.
We consider Lagrangian perturbations so that $\delta M=0$.
For each pressure component we have $P_i=\kpi\rho^\gpi$.
Prior to the perturbation, $\kpi$ is constant in space; after
the perturbation, $\kpi$ will generally have a different value,
and it may depend on position.
In general, we find
\beq
\delta\ln P_i=\delta\ln \kpi-\gpi\delta\indep,
\label{eq:delta_ln_P_i}
\eeq
where we have used the fact that $\rhoc$ is defined to be
the central density of the unperturbed sphere, so that
$\delta\ln\rho=-\delta\indep$.
The perturbation in the total pressure is then
\beq
\delta \ln P=\sum_i\pip\delta\ln P_i=\sum_i\pip\dlk-\gp\delta\indep.
\label{eq:dlnP}
\eeq
We can evaluate $\delta\indep$ 
from the fact that, at constant mass,
the definition of $\mu$ implies that $\mu\propto \rho^2P^{-3/2}$.
Varying this relation and combining it
with equation (\ref{eq:dlnP}), we find
\beq
\delta\indep=-\frac{2}{4-3\gp}\left[\dlm+\frac32\sum_i\pip\dlk\right],
\label{eq:deltaxi}
\eeq
and
\beq
\delta\ln P=\frac{2}{4-3\gp}\left[\gp\dlm+2\sum_i\pip\dlk\right].
\label{eq:dlnp1}
\eeq

	We now apply Lagrangian variations to the structure equations.
Starting with equation (\ref{eq:d_ln_M_2}), we have
\beq
{d\delta\indep\over d\indep}=-{\delta\gp\over\gp}-
	\delta\ln\left(\lambda^4\over\mu^2\right).
\label{eq:ddeltaxi}
\eeq
It is convenient to
insert equation (\ref{eq:d_ln_M_2}) into equation (\ref{eq:mu_diff})
before carrying out its variation:
\beq
d\ln\mu=d\ln M-\frac12(4-3\gp)d\indep.
\label{eq:d_ln_mu}
\eeq
Varying equations (\ref{eq:d_ln_mu}) and (\ref{eq:lambda_diff}), and
using equation (\ref{eq:ddeltaxi}), we obtain the equations
for $d\dlm/d\chi$
(eq. \ref{eq:ddlm}) and $d\dll/d\chi$
(\ref{eq:ddll}) in the text.

\subsection{Evaluation of $\delta\gp$ for adiabatic perturbations}

	To complete the set of equations for the perturbed
structure, we must determine 
the perturbation in the polytropic index, $\delta\gp$.
Recall that a perturbed locally adiabatic polytrope is not
itself a polytrope unless it is isentropic (\S2.2).
As a result, we must allow for a variation in $\kpi$ when
we evaluate the polytropic index:
\beq
\gp\equiv{d\ln P\over d\ln\rho}=
\sum_i \pip\left(\gpi-{d\ln \kpi\over d\indep}\right),
\label{eq:gamma_p_III}
\eeq
from equation (\ref{eq:delta_ln_P_i}).
(In equilibrium, the cloud is a polytrope,
$\kpi$ is independent of position, and
$\gp$ reduces to $\sum\gpi(P_i/P)$ as it 
must---see eq.~[\ref{eq:gpsum}].)
Varying equation (\ref{eq:gamma_p_III}), we obtain
\beq
\delta\gp=\sum_i\pip\left[(\gpi-\gp)\dlk-\gpi(\gpi-\gp)\delta\indep
	-{d\dlk\over d\indep}\right],
\label{eq:delta_gp}
\eeq
with the aid of equation (\ref{eq:delta_ln_P_i}).
We have used the fact that  $\kpi$ is constant
($d\ln \kpi/d\indep=0$) in the initial equilibrium state
in order to simplify this result.

	For a locally adiabatic pressure component $\ell$, we have 
$P_\ell=K_\ell\rho^\gl$, where the entropy parameter $K_\ell$
is constant during a perturbation.  As a result, we have
$\delta\ln P_\ell=\gl\delta\ln\rho=-\gamma_\ell\delta\indep$, so that
\beq
\delta\ln \kpl=(\gpl-\gl)\delta\indep,
\label{eq:lnkpl}
\eeq
from equation (\ref{eq:delta_ln_P_i}).
As a result, such a component contributes a term
\beq
(\delta\gp)_\ell=\plp
\left[(\gamma_\ell-\gpl){d\delta\indep\over d\indep}-
\gamma_l(\gpl-\gp)\delta\indep\right]
\label{eq:locally_delta_gp_l}
\eeq
to $\delta\gp$.

	For a globally adiabatic pressure component, it is
the total entropy of the cloud that remains constant during
a perturbation.  In order to be distinct
from the locally adiabatic case, the component must be non-isentropic
($\gamma_g\neq\gpg$).
Redistribution of energy in the cloud during the perturbation
maintains the polytropic form,
$P_g\propto\rho^{\gpg}$,
but changes the value of $\kpg$ (the entropy parameter $K_g$
changes as well).
The change in $\kpg$ needed
to maintain constant entropy will be discussed in \S6 below.
All that we need to note here is that since
$\kpg$ is independent of position for a globally adiabatic
pressure component, such a component contributes a term
\beq
(\delta\gp)_g={P_g\over P}(\gpg-\gp)
	(\delta\ln \kpg-\gpg\delta\indep)
	\label{eq:globally_delta_gp_g}
\eeq
to $\delta\gp$ from equation (\ref{eq:delta_gp}).

	In a multi--pressure polytrope, both types of pressure
components may be present.  
For simplicity, we shall assume that only one globally
adiabatic component is present; the generalization to more than
one such component is straightforward.
We find it convenient to introduce a new adiabatic index
\beq
\Gamma\equiv\sum_\ell \pip\gamma_i + \pgp\gpg,
\eeq
which is the weighted mean of the adiabatic indexes for the locally
adiabatic components $\ell$ and the polytropic index for the globally
adiabatic component $g$.  This is equivalent to
\beq
\Gamma=\gp+\sum_{\ell}\pip(\gamma_i-\gpi).
\label{eq:Gamma}
\eeq
A related quantity that we shall need is
\beq
\Gamma'\equiv\sum_{i\in l}\pip\gamma_i(\gp-\gpi) + 
	\pgp\gpg(\gp-\gpg).
\eeq
In terms of these quantities, we find
\beq
\delta\indep = -\frac{2}{4-3\Gamma}\left[\dlm+\frac32\pgp\dlkg\right],
\label{eq:deltaxi2}
\eeq
from equations (\ref{eq:deltaxi}) and (\ref{eq:dlnp1}). 

	Evaluating 
$\delta\ln\gp$ from equation (\ref{eq:delta_gp}), we then obtain
equation (\ref{eq:delta_ln_gp}) in the text.

\section{ENTROPY FOR GLOBALLY ADIABATIC PRESSURE COMPONENTS}
\label{app:global}

\subsection{Determination of the Entropy}

	Our objective is to evaluate the entropy parameter $\kgr$
(eq. \ref{eq:kgr})
for a globally adiabatic pressure component $g$ for
an arbitrary non-uniform cloud.
To do this, we generalize McKee \& Zweibel's
(1995) treatment of Alfv\'en waves to arbitrary values of
$\gamg$ and $\gpg$.  The equation for the internal energy of
an ideal gas is
\beq
\frac{\partial u_{g}}{\partial t}+{\bf\nabla}\cdot(u_{g}{\bf v}+
	{\bf q}_{g})+P_{\rm g}{\bf \nabla\cdot v}={\cal S}_{g},
\label{eq:then}
\eeq
where ${\bf q}_{g}$ is the heat flux
and ${\cal S}_{g}$ is the source term for the globally adiabatic
pressure component (e.g., McKee et al 1987).
(Note that it is possible for ${\cal S}_{g}$ to vanish in our model even in
the presence of heating and cooling, provided the effects of
the heating and cooling are represented by internal degrees of
freedom.)
The total internal energy in the cloud 
associated with globally adiabatic pressure component
is $U_{g}=\int u_{g} dV$,
where the integral extends over the entire cloud.
The rate of change of this energy is
\beq
\frac{dU_{g}}{dt}  = \int \frac{\partial u_{g}}{\partial t}\; dV
	+\int u_{g}{\bf v\cdot dS}.
\eeq
Eliminating $\partial u_{g}/\partial t$ with equation (\ref{eq:then})
and using the divergence theorem, this becomes
\beq
\frac{dU_{g}}{dt}  = \int[-P_{g}{\bf\nabla\cdot v}+
	{\cal S}_{g}]\,dV-\int{\bf q}_{g}\bf{\cdot dS},
\label{eq:dudt1}
\eeq
where ${\bf dS}$ is an element of the surface bounding the system.
For an adiabatic process, the last two terms vanish.  Note that
we do not assume that the heat flux is zero inside the cloud,
only at the surface.  
We do assume 
that there is no exchange of heat with the other pressure components;
for example, we do not allow for the possibility that energy would
be transferred from the waves to the thermal motions of the
gas by wave damping.
The equation of continuity then allows us
to rewrite equation (\ref{eq:dudt1}) as
\beq
\frac{dU_{g}}{dt}=\int P_{g}\,\frac{d\ln\rho}{dt}\,dV 
	     =\kpg(t)\int\rho^{\gpg-2}\frac{d\rho}{dt}\,dM,
\label{eq:dudt2}
\eeq
where we have included the argument in $\kpg(t)$ to emphasize
that it may change with time.
Since the internal energy is given by
\beq
U_{g}=\frac{1}{\gamg-1}\int P_{g} dV=
	\frac{\kpg}{\gamg-1}\int\rho^{\gpg-1} dM,
\label{eq:u1}
\eeq
we can eliminate the unknown $\kpg$ from equation (\ref{eq:dudt2}),
\beq
\frac{d\ln U_{g}}{dt}=\left(\frac{\gamg-1}{\gpg-1}\right)\frac{d}{dt}\;
	\ln\left(\int\rho^{\gpg-1}dM\right).
\label{eq:dlnu}
\eeq
The internal energy can also be expressed in terms of the mean square
sound speed $\cavgt$,
\beq
U_{g}= \frac{1}{\gamg-1} \int \left(\frac{P_{g}}{\rho}\right) 
	dM \equiv \frac{1}{\gamg-1} M\cavgt.
\eeq
As a result, equation ({\ref{eq:dlnu}}) can be re-expressed as
\beq
\frac{d\ln\cavgt}{dt}=\left(\frac{\gamg-1}{\gpg-1}\right)\frac{d}{dt}\;
	\ln\left(\frac{\cavgt}{\kpg}\right).
\eeq
Integrating this and evaluating the integration constant
in the reference state, we find
\beq
\kgr^{\gpg-1}=\kpg^{\gamg-1}\cavgt^{-(\gamg-\gpg)},
\label{eq:kgr2}
\eeq
which is the desired result.  

	Next, we express the entropy parameter for a globally
adiabatic component in a form similar to 
that for a locally adiabatic component, $\avg{K_\ell}$.
We define the mean entropy parameter as
\beq
\kavg^{\left(\frac{\gpg-1}{\gpg-\gamg}\right)}\equiv\frac{1}{M}\int
	K_g^{\left(\frac{\gpg-1}{\gpg-\gamg}\right)} dM.
\label{eq:kavgb}
\eeq
Note that in order to keep the notation simple, we have
deviated from our convention of using $\avg{x}$ to
mean $M^{-1}\int x\, dM$; we made a similar exception
for $\kavl$.  
From dimensional analysis (see eq. \ref{eq:kgr2})
or by direct evaluation, we obtain
\beq
K_g^{\gpg-1}=\kpg^{\gamg-1}c_g^{-2(\gamg-\gpg)}.
\eeq
Inserting this into equation (\ref{eq:kavgb}),
we find that $\kavg$ is just
the entropy in the reference state,
\beq
\kavg=\kgr.
\eeq

	Equation (\ref{eq:kgr2}) relates the  
entropy parameter $\kgr=\kavg$ to the polytropic parameter $\kpg$
in terms of $\cavgt$, which is proportional
to the mean temperature of the pressure component.  In the particular case
of an isentropic component, we have simply $\kavg=\kpg$, 
as was clear from the definitions.
For the globally adiabatic case of greatest
interest to us, Alfv\'en waves, we have $\gamg=3/2$ and $\gpg=1/2$, so 
that $\kavg=\cavgt^2/\kpg$.
With the aid of equation (\ref{eq:kgr2}), we can express
the entropy parameter in terms of directly observable quantities
if we choose the pressure in the
reference state to be the surface pressure $P_{\rm gs}$:
\beq
\kavg=\frac{c_{g{\rm s}}^{2\gamg}}{\psi_g^{\left({\gamg-\gpg\over
	\gpg-1}\right)}P_{g{\rm s}}^{(\gamg-1)}},
\label{eq:kref}
\eeq
where $\psi_g\equiv\cavgt/c_{g\s}^2$ is usually of order unity,
and we have included the subscript ``s" to emphasize
that the quantity is to be evaluated at the surface of the cloud.

\subsection{Evaluation of $\dlkg$ at Constant Entropy}

	To evaluate $(\delta\gp)_{g}$ in equation 
(\ref{eq:globally_delta_gp_g}),
we need to know how $\kpg$ changes as the 
result of a globally adiabatic perturbation.  Variation of
equation (\ref{eq:kgr2}) at constant entropy gives
\beq
(\gamma_{g}-1)\dlkg=(\gamg-\gpg)\delta\ln\cavgt.
\label{eq:dlk1}
\eeq
Now, with the aid of equation (\ref{eq:delta_ln_P_i}) we find
\beq
\delta c_{g}^2=\frac{P_{g}}{\rho}(\delta\ln P_{g}-\delta\ln\rho)
	=c_{g}^2\left[\dlkg-(\gpg-1)\delta\xi\right].
\eeq
Noting that $\dlkg$ is constant inside the cloud, we obtain
\beq
\delta\ln\cavgt=\dlkg-\frac{\gpg-1}{M\cavgt}\int c_{g}^2\delta\xi dM.
\eeq
Inserting this result into equation (\ref{eq:dlk1}), we then find
\beq
\dlkg={\displaystyle
	2(\gamma_{g}-\gpg)\int\left(\frac{\dlm}{4-3\Gamma}\right)
	c_{g}^2 dM
	\over \displaystyle
	\int\left(\frac{4-3\sum\gi(P_i/P)}{4-3\Gamma}\right)
	c_{g}^2 dM}.
\label{eq:dlk2}
\eeq
with the aid of equation (\ref{eq:deltaxi2}) for $\delta\xi$.

\newpage

\figcaption[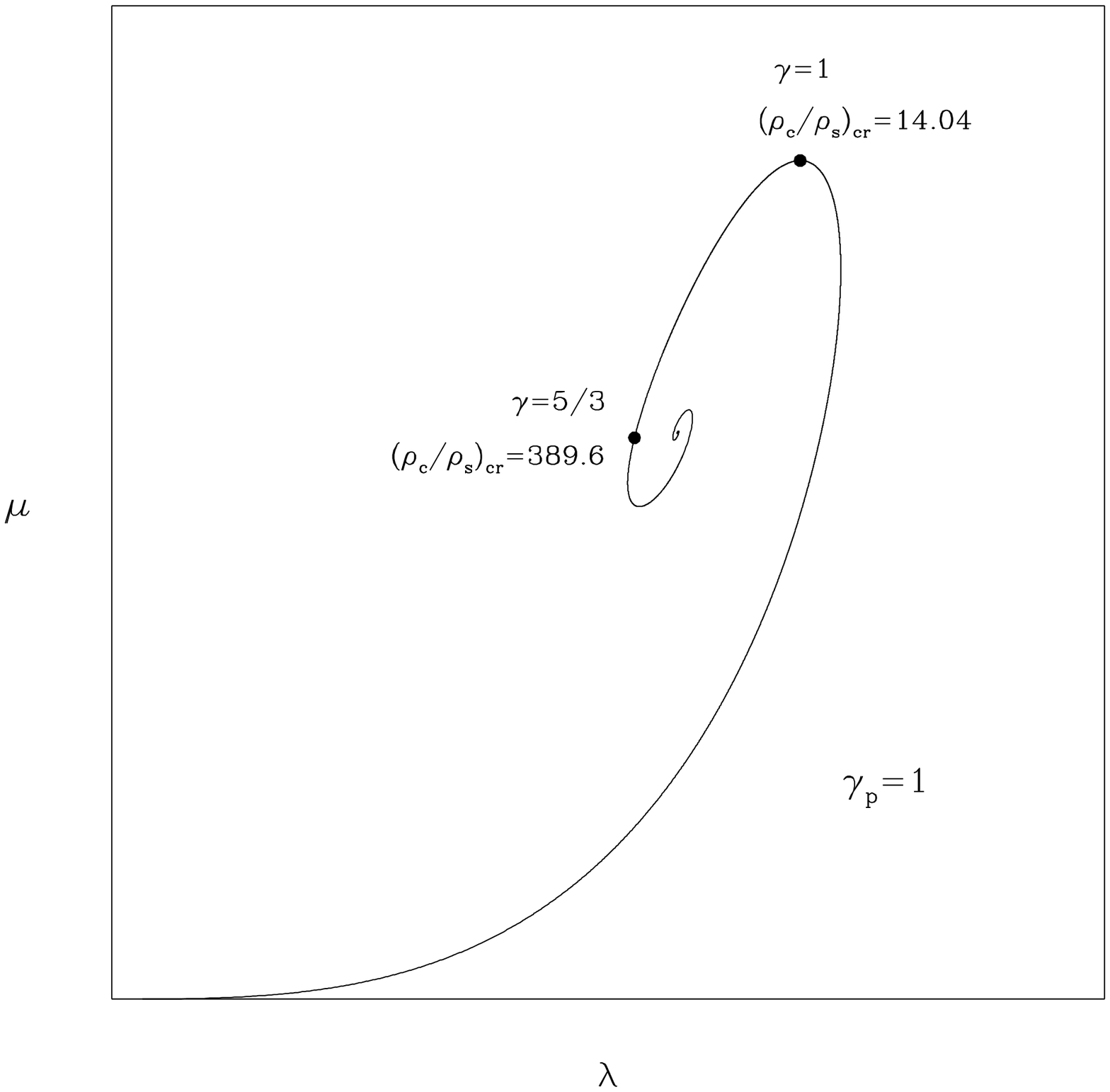]{Structure of a spatially isothermal
polytrope ($\gp=1$) in the $\mu-\lambda$ plane.  The critical
density ratios for an isentropic sphere ($\gamma=1$) and 
a globally adiabatic sphere with $\gamma=5/3$ are indicated.
\label{fig:ml}}

\figcaption[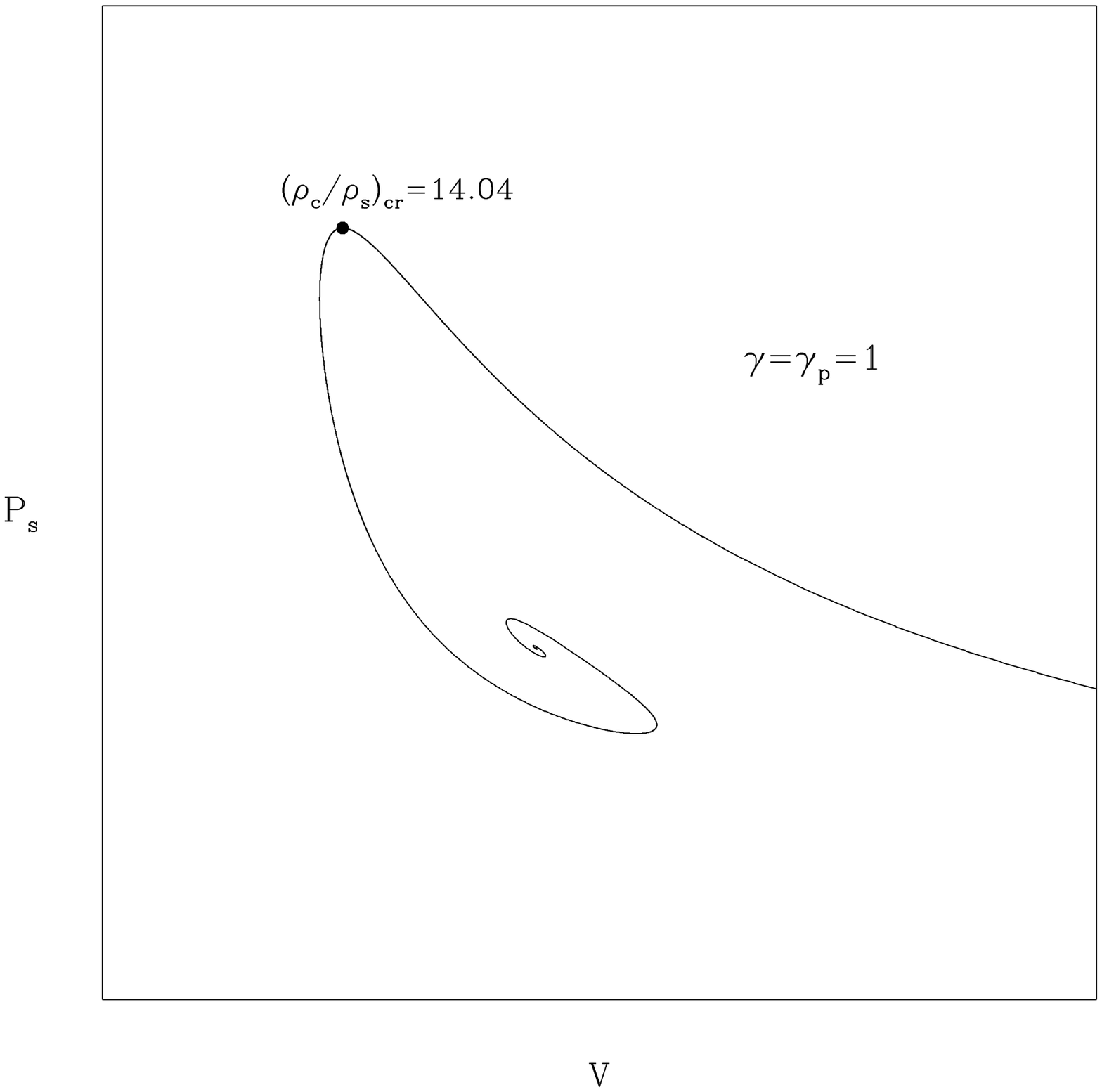]{Structure of a spatially isothermal
polytrope ($\gp=1$) in the $P_\s-V$ plane.  The critical point
for $\gamma=1$ occurs at the maximum value of the surface pressure $P_\s$.
\label{fig:pv}}

\figcaption[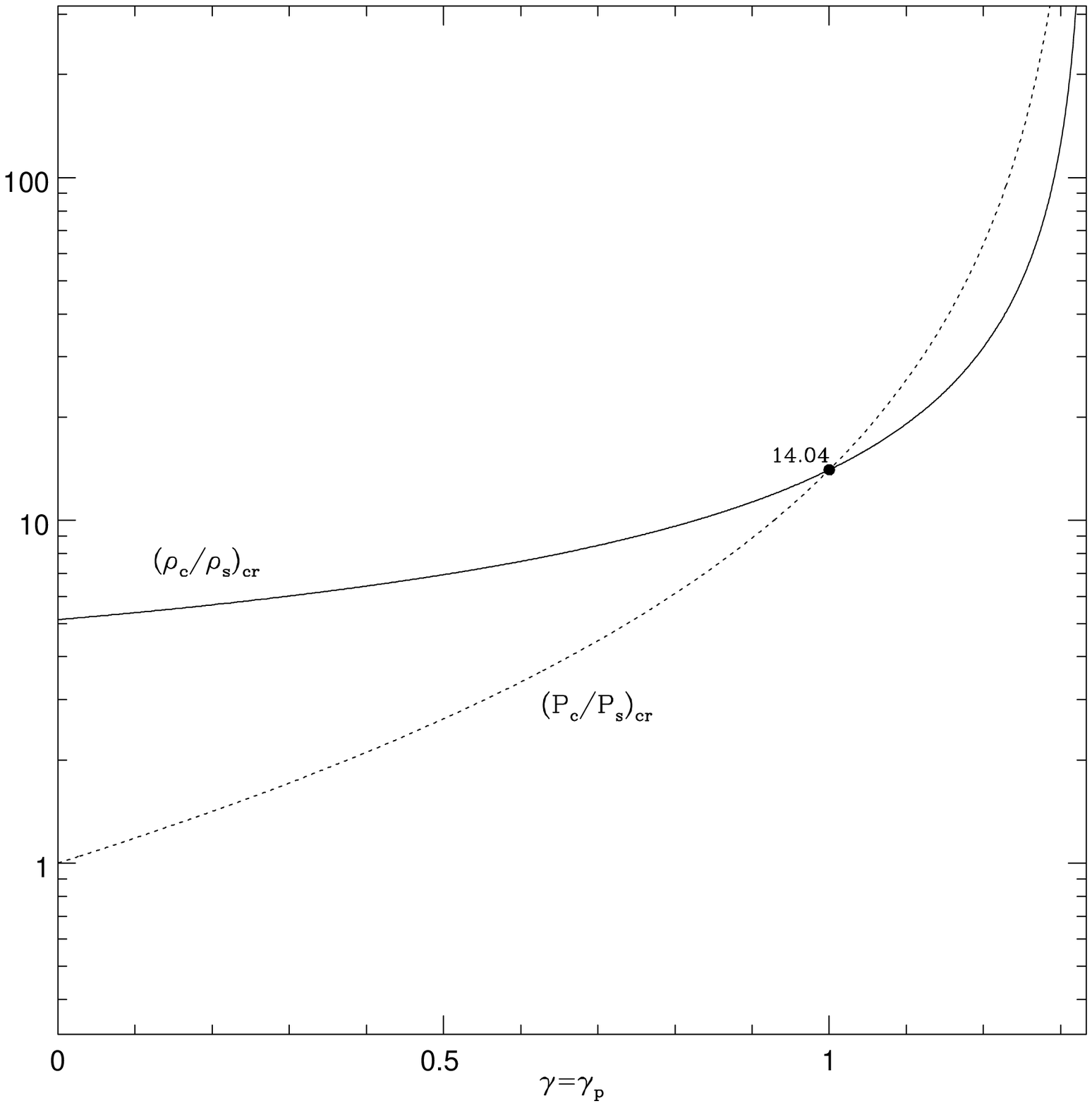]{The central--to--surface density
and pressure ratios for isentropic polytropes ($\gamma=\gp$).
The numerical value is for the isothermal case.
\label{fig:rho}}

\figcaption[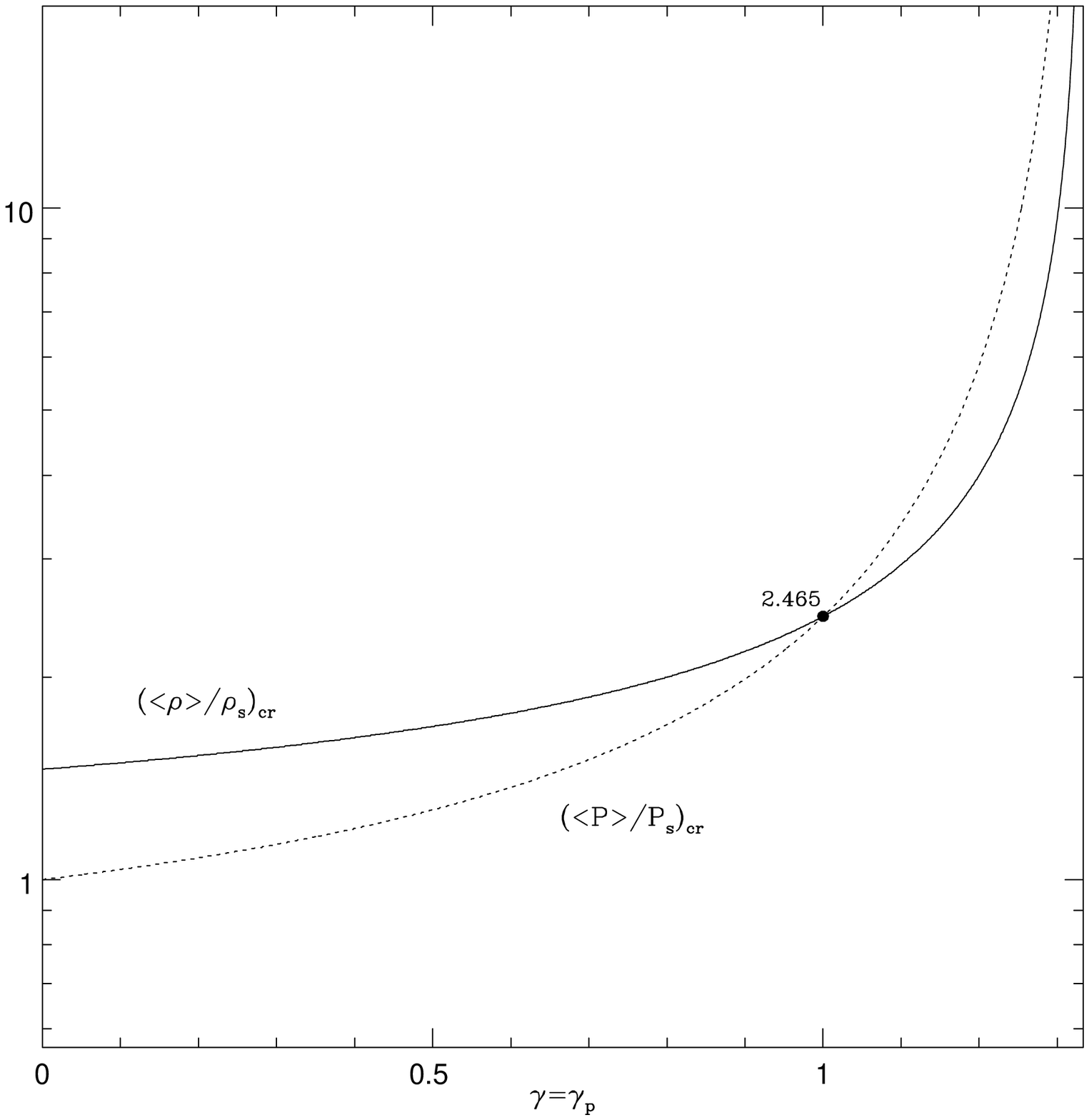]{The mean--to--surface density
and pressure ratios for isentropic polytropes ($\gamma=\gp$).
The numerical value is for the isothermal case.
\label{fig:rhobar}}

\figcaption[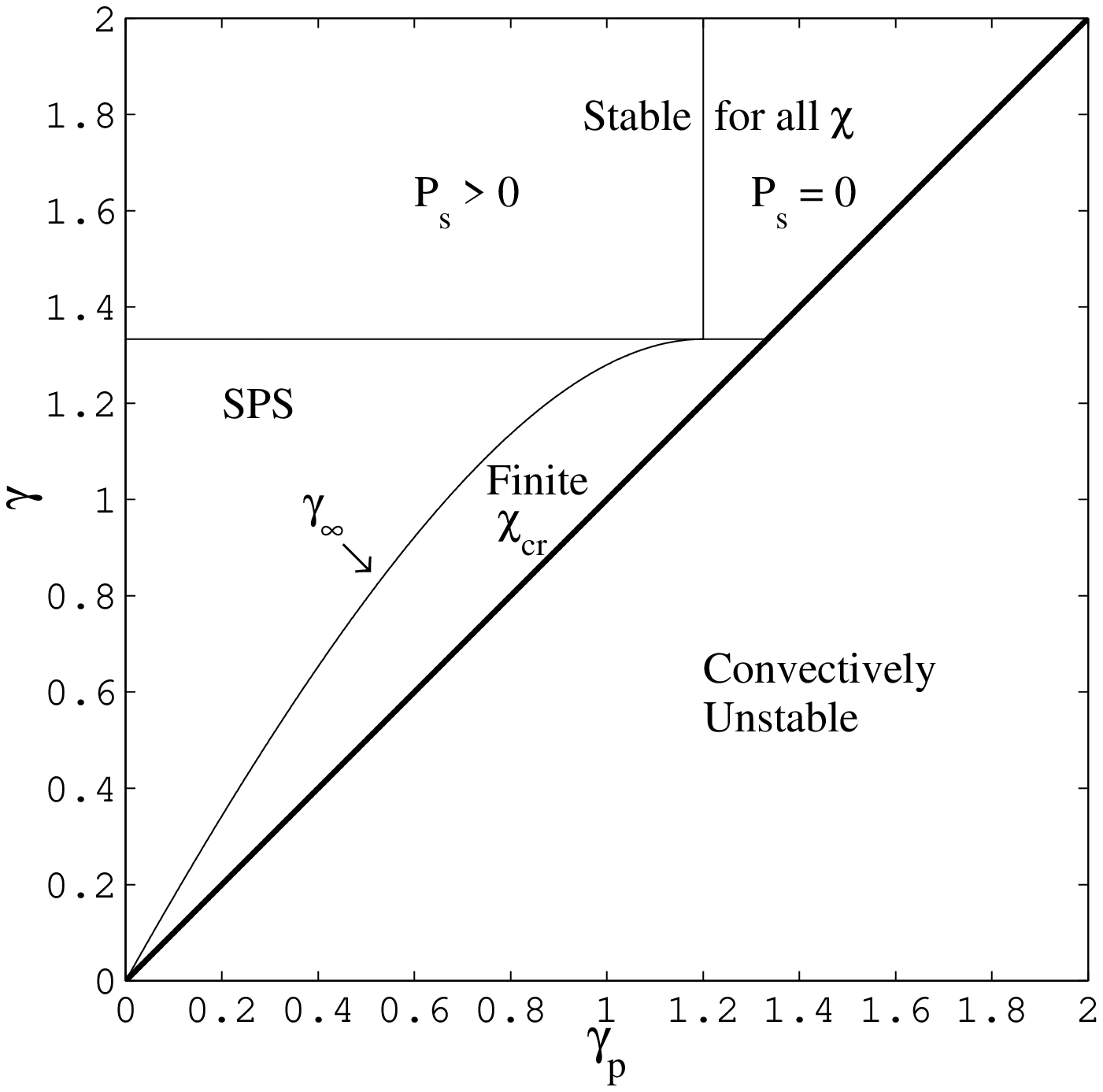]{Characteristics of locally adiabatic
polytropes in the $\gamma-\gp$ plane.  We do not consider
polytropes in the region $\gamma<\gp$ since they are convectively
unstable.  Above $\gamma=4/3$, the polytropes are always
stable, even for arbitrarily large values of
$\chi=\ln(\rho_{\rm c}/\rho_\s)$. 
Stellar polytropes have $\gamma>4/3$ so that they
are stable, and in addition have $\gp>6/5$ so that
they can have $P_\s=0$ at finite radius; as a result,
they can be used as models for stable stars.
Stable, finite--sized polytropes in the rest
of the parameter space with $\gamma\geq\gp$ must be confined by the
pressure of an ambient medium.
The locally adiabatic
pressure components in interstellar
clouds  have $\gi\leq 4/3$.
For polytropes in the region
labelled ``Finite $\chi_{\rm cr}$", the polytropes become
unstable for a central--to--surface density contrast
greater than $\exp(\chi_{\rm cr})$.  In the region
labelled ``SPS" (singular polytropic sphere), 
polytropes are formally stable
even for an infinite density contrast ($\chi=\infty$).
These regions are separated by the curve labelled
$\gamma_\infty$, which is given in equation (\ref{eq:ginfty}). 
It is possible that clouds in part of the SPS region,
particularly near the $\gamma_\infty$ curve
at small $\gp$, are unstable to finite perturbations (\S 
\ref{sec:pressure}).
\label{fig:laggb}}

\figcaption[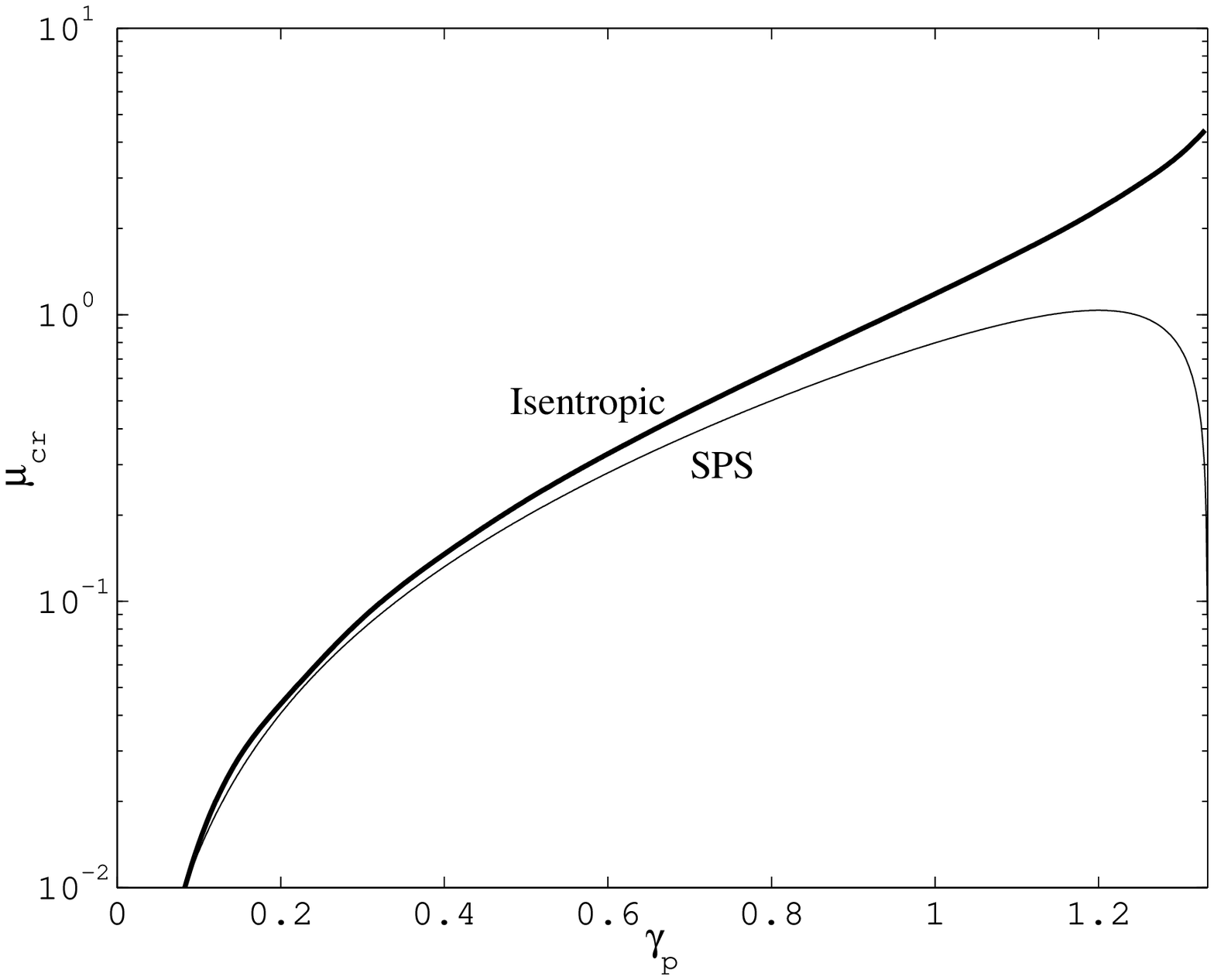]{The value of $\mucr$, which determines
the critical mass, is plotted for isentropic polytropes
as a function of $\gp$.
Also shown is the value of $\mu$ for singular
polytropic spheres (labeled SPS).  For polytropes with
a critical point, we have $\mucr$(isentropic)$\geq\mucr
\simgt\mu$(SPS).
\label{fig:mucr}}

\figcaption[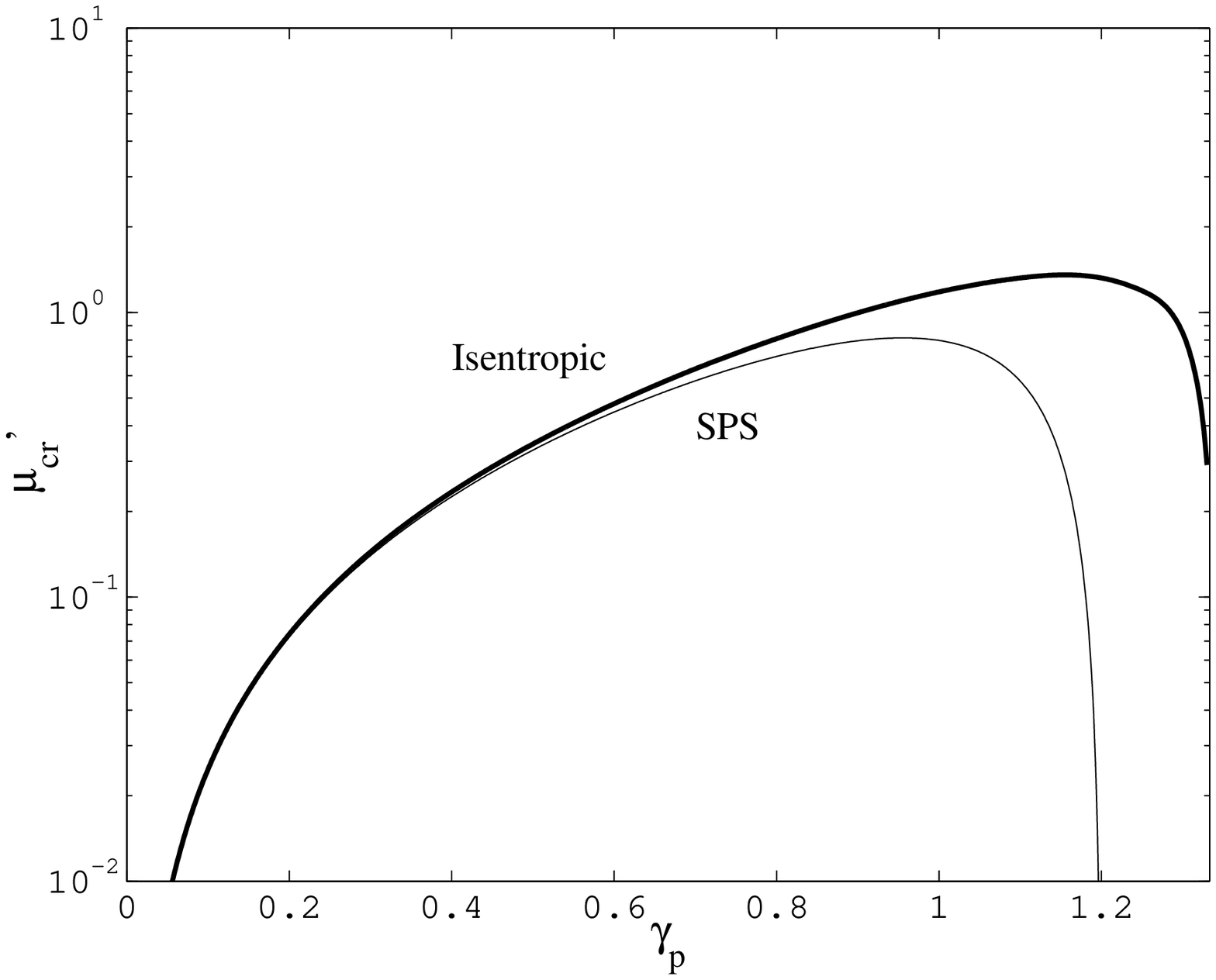]{The same as Figure \ref{fig:mucr},
except that the dimensionless critical mass of the isentropic
polytrope and the dimensionless mass of the singular polytropic
sphere are given in terms of the rms
value of $c^2$ instead of the surface value.
\label{fig:mucrp}}

\figcaption[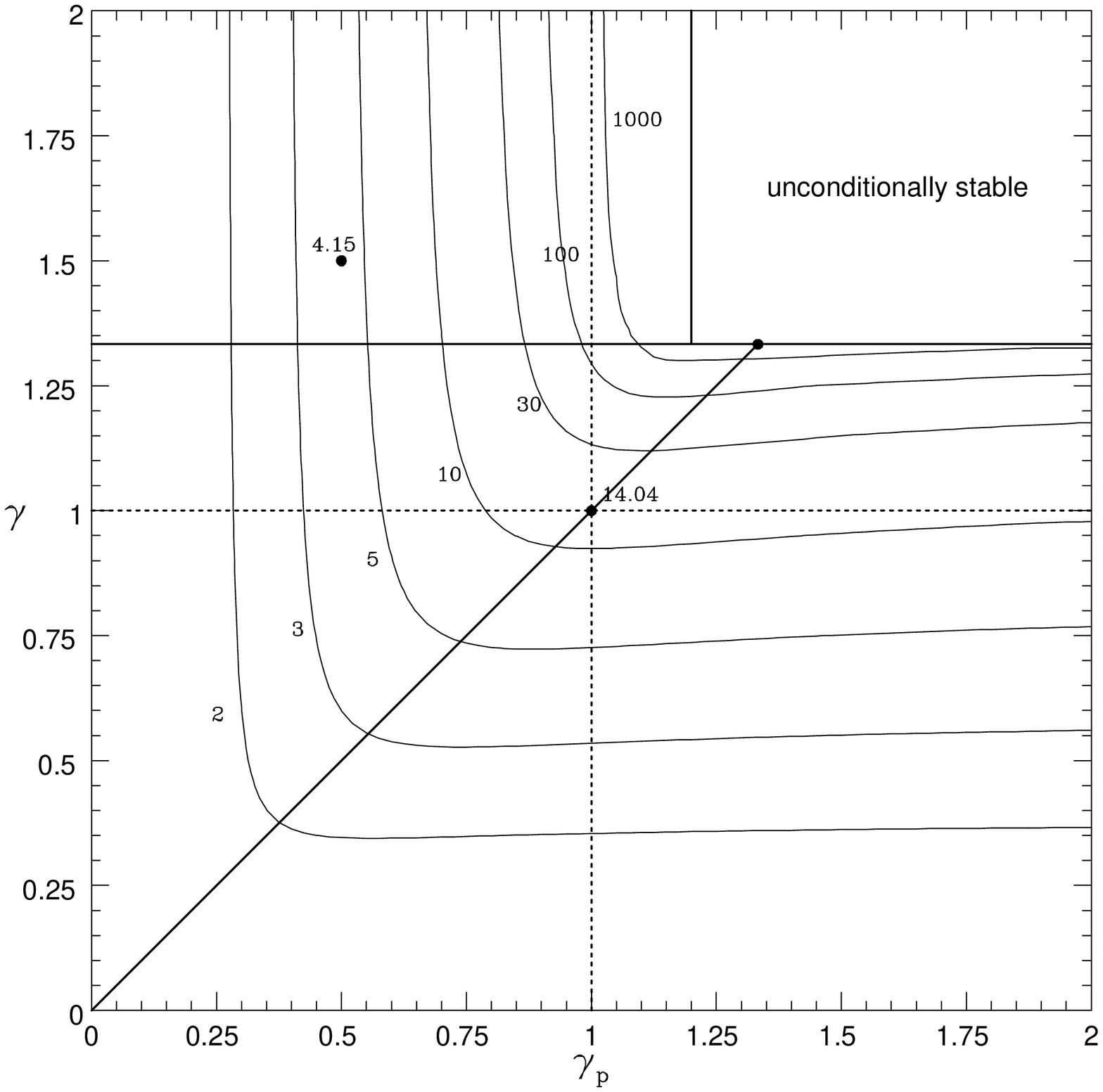]{The central--to--surface
pressure ratio for globally adiabatic polytropes at
the critical point.
Since such polytropes cannot be convectively unstable,
it is possible to have stable clouds with $\gamma<\gp$.
Globally adiabatic polytropes with $\gamma>4/3$ are unstable
to core collapse for $\gp<6/5$.  The point at $\gp=0.5$
and $\gamma=3/2$ is appropriate for small amplitude
Alfv\'en waves.
\label{fig:gapr}}

\figcaption[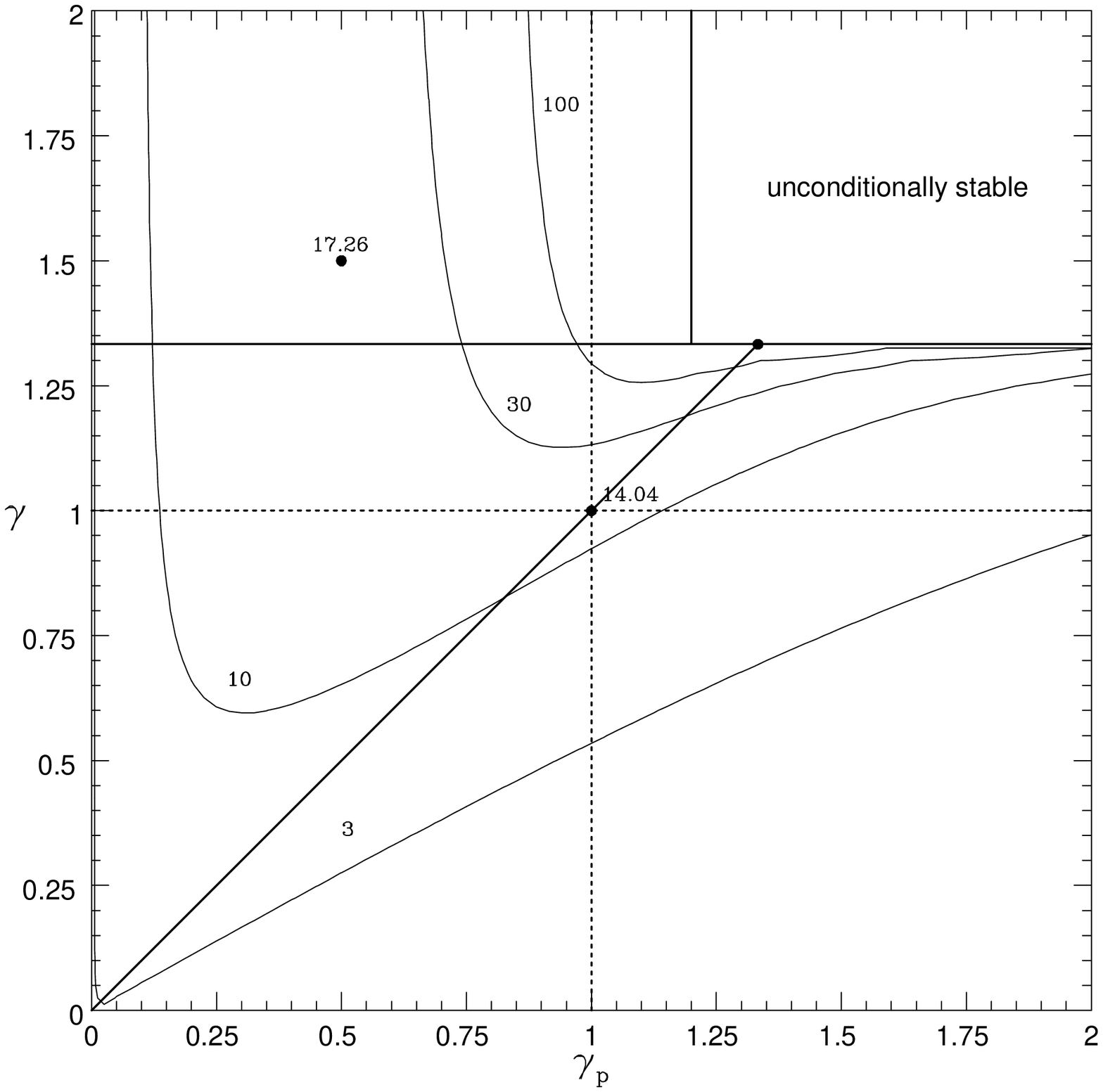]{The same as Figure \ref{fig:gapr},
except that the curves represent the central--to--surface
density ratio at the critical point. 
\label{fig:gaden}}

\figcaption[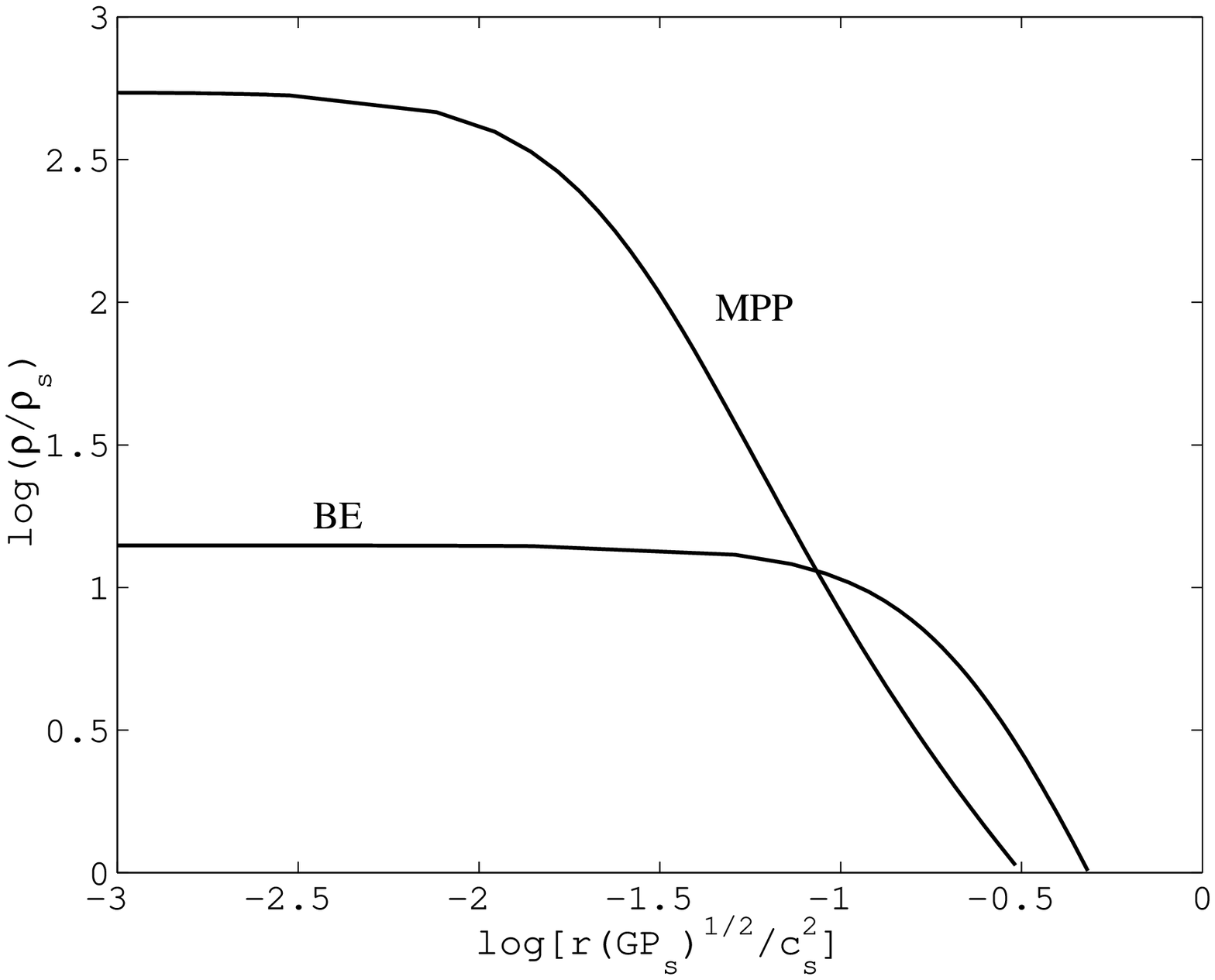]{The density distribution
of a multi--pressure polytrope that has
$P_B\simeq P_{\rm w}\simeq 10\pth$,
similar to the values observed by Crutcher (1999)
in a sample of molecular clouds.  
The density is normalized to the density at the cloud
surface, $\rho_\s$, and the radius is normalized to the
value of $(c_\s^2/GP_\s)^{1/2}=c_\s/(G\rho_\s)^{1/2}$
at the cloud surface; thus, at the surface, the abscissa $=\lambda$.  
We have assumed that
$\gamma_{{\rm p}, B}=1.0$ and that the cloud is close
to, but not at, the critical state.  For comparison, 
we have portrayed the density distribution of a
Bonnor--Ebert sphere with the same surface pressure
and density.  The multi--pressure polytrope has a much
larger density drop than the Bonnor--Ebert sphere because
of the stabilizing effects of the non--isentropic magnetic pressure;
the mean--to--surface density drops for the two cases are
about the same, however.
\label{fig:mpp}}

\end{document}